\documentclass[12pt]{article}

\usepackage{setspace}

\usepackage{graphicx}
\usepackage{epstopdf}
\usepackage[body={17.5cm, 21cm},right=2cm]{geometry}
\usepackage{amssymb}
\usepackage{amsmath}
\usepackage{dcolumn}

\usepackage{graphicx}
\usepackage{epstopdf}
\usepackage{amsmath}
\usepackage{amsfonts}
\usepackage{amssymb}
\usepackage[usenames]{color}
\usepackage{mathrsfs}
\usepackage{array}

\usepackage{graphicx}
\usepackage{dcolumn}
\usepackage{bm}
\usepackage{amssymb}
\usepackage{epstopdf}
\usepackage{amsmath}
\usepackage{amsfonts}
\usepackage{color}
\usepackage{mathrsfs}
\usepackage{comment}

\usepackage{multirow}
\usepackage{psfrag}
\pdfoutput=1

\usepackage[colorlinks,bookmarks]{hyperref}
\definecolor{linkblue}{rgb}{0,0,0.8}
\definecolor{linkgreen}{rgb}{0,0.5,0}

\hypersetup{pdfpagemode=UseNone, pdfstartview=FitH, linkcolor=linkblue, %
            citecolor=linkgreen, urlcolor=linkblue}
            
\usepackage[numbers,sort&compress]{natbib}

\newcommand{\be}{\begin{equation}}
\newcommand{\ee}{\end{equation}}
\newcommand{\bea}{\begin{eqnarray}}
\newcommand{\eea}{\end{eqnarray}}
\newcommand{\barr}{\begin{array}}
\newcommand{\earr}{\end{array}}

\newcommand{\thetabubble}{\theta_{\rm bubble}}

\def\tzeta{{\tilde\zeta}}
\def\n{{\bf \hat n}}

\def\k{{\bf k}}
\def\x{{\bf x}}
\def\tzeta{{\tilde\zeta}}
\def\hk{{\bf \hat k}}
\def\bigoh{{\mathcal O}}
\def\E{{\mathcal E}}
\def\beq{\begin{equation}}
\def\eeq{\end{equation}}
\def\be{\begin{equation}}
\def\ee{\end{equation}}
\def\bea{\begin{eqnarray}}
\def\eea{\end{eqnarray}}

\def\nn{\nonumber}
\def\n{{\bf \widehat n}}
\def\k{{\bf k}}
\def\ba{\begin{align}}
\def\ea{\end{align}}

\def\L{{\mathcal L}}
\def\Nside{N_{\rm side}}
\def\Npix{N_{\rm pix}}

\newcommand\lsim{\mathrel{\rlap{\lower4pt\hbox{\hskip1pt$\sim$}}
        \raise1pt\hbox{$<$}}}
\newcommand\gsim{\mathrel{\rlap{\lower4pt\hbox{\hskip1pt$\sim$}}
        \raise1pt\hbox{$>$}}}

\begin{document}

\vspace{5mm}
\vspace{0.5cm}
\begin{center}

\def\thefootnote{\fnsymbol{footnote}}

{\Large \bf Optimal analysis of azimuthal features in the CMB}
\\[0.8cm]

{\large Stephen J.~Osborne$^{1}$, Leonardo Senatore$^{1,2,3,4}$ and Kendrick Smith$^{4,5}$}
\\[0.5cm]

{\normalsize { \sl $^{1}$ Department of Physics, Stanford University, Stanford, CA 94306}}\\
\vspace{.3cm}

{\normalsize { \sl $^{2}$ Kavli Institute for Particle Astrophysics and Cosmology, \\ Stanford University and SLAC, Menlo Park, CA 94025}}\\
\vspace{.3cm}

{\normalsize { \sl $^{3}$ Stanford Institute for Theoretical Physics and Department of Physics, \\Stanford University, Stanford, CA 94306}}\\
\vspace{.3cm}

{\normalsize { \sl $^{4}$ CERN, Theory Division, 1211 Geneva 23, Switzerland}}\\
\vspace{.3cm}

{\normalsize { \sl $^{5}$ Department of Astrophysical Sciences, Princeton University, Princeton, NJ 08544-1001, USA}}\\
\vspace{.3cm}

{\normalsize { \sl $^{6}$ Perimeter Institute for Theoretical Physics, Waterloo, ON N2L 2Y5, Canada}}\\
\vspace{.3cm}

\end{center}

\vspace{.8cm}

\hrule \vspace{0.3cm}
{\small  \noindent \textbf{Abstract} \\[0.3cm]
We present algorithms for searching for azimuthally symmetric features in CMB data.
Our algorithms are fully optimal for masked all-sky data with inhomogeneous noise,
computationally fast, simple to implement, and make no approximations.
We show how to implement the optimal analysis in both Bayesian and frequentist cases.
In the Bayesian case, our algorithm for evaluating the posterior likelihood is so
fast that we can do a brute-force search over parameter space, rather than using
a Monte Carlo Markov chain.
Our motivating example is searching for bubble collisions, a pre-inflationary signal
which can be generated if multiple tunneling events occur in an eternally inflating
spacetime, but our algorithms are general and should be useful in other contexts.
\noindent 
}
 \vspace{0.3cm}
\hrule
\def\thefootnote{\arabic{footnote}}
\setcounter{footnote}{0}

\clearpage

\section{Introduction}


What happened at the beginning of the universe? How did the universe start? Are there other
universes? What do they look like?
Obtaining answers to these exciting questions is theoretically and experimentally challenging,
but there are known signals that may be present in the Cosmic Microwave Background (CMB)
data that can help us to answer them.

One way to solve the horizon, flatness, and relic abundance problems is if
the universe started with a phase of slow roll inflation.
There are theorems~\cite{Borde:2001nh} that show that inflation cannot be past eternal:
there must be something before it.
One possibility is that our current patch of universe was born as a nucleation bubble
from a phase of false vacuum eternal inflation~\cite{Guth:1980zm,Guth:1982pn}.
In this phase the universe is thought to be trapped in an unstable high-energy vacuum. The vacuum
energy makes the universe expand exponentially, but since the vacuum is quantum mechanically unstable
to tunneling, new bubbles are continuously produced and start expanding at close to the speed of light.
If the decay rate per unit spacetime volume is less than $H^4$, with $H$ the Hubble rate, then
the expansion of the universe draws the bubbles far apart and they do not fill the universe.
Inside the bubble the universe looks like an open FRW cosmology, with a big bang apparent singularity
at the spacetime location of the nucleation. Inflation can occur within the bubble and
produce a universe that looks locally like our own. In this scenario, the singularity in the past before
inflation is an illusion, and we came from an eternally inflating space-time.

Though bubbles do not percolate and fill the whole of space, there is a chance that bubbles
collided before our present time, leaving a specific disk-shaped imprint in the CMB.
Discovery of such a signature would have consequences that can hardly be overstated.
First, we would better understand what happened before the period of inflation in our recent
past. We would learn that we are a bubble in an eternally inflating universe, and that eternal inflation
consists of a new phase of our universe.
Furthermore, detection of a bubble would provide indirect confirmation of the anthropic explanation of the
cosmological constant~\cite{Weinberg:1987dv}, which states that the observed value of the cosmological
constant is approximately the value required for structures to form in
our universe. For structure formation the cosmological constant cannot be larger than a certain
upper bound, and since it is most probable for the cosmological constant to be peaked at the highest
possible value, then it follows that we should observe a cosmological constant close to its
upper bound. And we did in 1998, confirming this prediction.
This anthropic explanation relies on the fact that the fundamental theory of the universe has a landscape
of vacua with different fundamental parameters, so that in one of them the right anthropic value
of the cosmological constant can be found. String theory naturally provides such a landscape of vacua, and
eternal inflation offers a way of populating them all in the universe.
While detection of a bubble collision in the CMB  would strictly speaking teach us that the universe is
described by a field theory with at least {\it two} vacua, one unstable and eternally inflating
and the other the stable terminal one, it would give evidence towards the anthropic explanation of the
cosmological constant, landscape of string theory, and ultimately string theory itself.

The bubble signal, if it exists, will likely be present at a low signal-to-noise ratio and so we will
require sophisticated algorithms to search for it.
The amplitude of the signal can be measured using statistically optimal estimators,
however, the large amount of data required, coupled with potentially complex noise properties
makes them computationally expensive to implement.
Approximate estimators can be used but this reduces the sensitivity to the signal, effectively
throwing away information.
We will discuss generally how optimal estimators can be implemented in a computationally efficient way
and use as an example the search for the bubble collision signal in full-sky CMB data.
Our approach is broadly applicable to many image processing problems, and
we will present the methodology in a general manner wherever possible.

There are several challenges in implementing the optimal analysis.
CMB maps have two components with very different statistical properties:
the instrument detector noise, often most easily described in real space,
and the CMB signal itself, which is more simply described in harmonic space.
Furthermore, the noise can be inhomogeneous and anisotropic.
The inhomogeneity means that the noise has a larger variance in some regions than others, and can be
caused by, for example, observing some areas of the sky for longer than other areas.
If the image has masked areas then the noise will be anisotropic, with pixels that are
completely masked described as having infinite noise variance.
It is difficult to avoid masking data in CMB analyses
since some foreground emission, such as from the galactic plane, is
many times brighter than the CMB at all frequencies.
The mask has both large and small scale features, and so algorithms
must account for the full range of scales when calculating the noise covariance.
Estimators constructed using matched filters, for example~\cite{McEwen:2006ke},
typically assume that the noise is isotropic, which is not the case in practice.
The resolution of the images also presents computational challenges.
To evaluate optimal estimators we must calculate the operation of the inverse signal+noise covariance matrix
on a vector. While it is simple to design algorithms that improve upon the
$\mathcal{O}(N_{\rm pix}^3)$ compute-time complexity of the simplest algorithms,
the compute times are still non-negligible even for more sophisticated algorithms
at the resolutions that we consider:
$\mathcal{O}(10^6)$ ($\mathcal{O}(10^8)$) pixels for the WMAP (Planck) experiment.
The image resolution is determined by the size of features in the signal, as well as the image
noise properties, and so can be decreased for large-scale signals.
An additional complication is the large number of parameter values that can be required to
describe the signal being searched for.
For example, the signal could be at any location on the sky, and have many possible angular profiles
that could be a complicated function of the model parameters.
These challenges appear to make calculating the
exact likelihood for all possible parameter values
very expensive.

In this paper, we present a complete solution to these computational problems.
We use a simple methodology, parametrizing the searched-for signal and calculating
the likelihood of the parameters, and finding algorithmic tricks to make the computations
fast.
We show how to implement the optimal analysis in both the Bayesian and frequentist
statistical frameworks.
The advantages of our method are simplicity, exactness, optimality, and minimal 
computational cost required.
For example, in the Bayesian case, we can compute the exact all-sky posterior likelihood
using a very straightforward procedure which is so computationally fast that we
can explore the parameter space by brute force, without MCMC-based sampling algorithms.

Previous searches have looked for the bubble signal in the WMAP data. 
Ref.~\cite{Feeney:2012hj} finds no evidence for bubble collisions, quoting an upper limit $N < 4.0$
on the expected number of collisions.
This analysis is based on exploration of the Bayesian likelihood and is similar in spirit to ours, but
there are a few minor differences.
First, our fast algorithms allow the likelihood analysis to be simplified while remaining computationally
affordable, thus removing several steps in the analysis.
Second, we evaluate the exact likelihood function which makes the analysis fully optimal (for example,
our filter is optimally weighted in the presence of sky cuts and inhomogeneous noise, and we can use
WMAP V-band data in addition to W-band).
Finally, we prefer to reparametrize and quote the final result as an upper limit on the maxmimum amplitude $A$ of a
bubble which intersects our Hubble volume, rather than an upper limit on the expected number of bubble collisions $N$.
The data analysis mainly constrains $A$, while the limit on $N$ is somewhat dependent on the prior on $A$ which is
chosen.
Notwithstanding these minor differences, we agree with the recent conclusion of~\cite{Feeney:2012hj}:
there is no statistically significant evidence for bubble collisions in WMAP.
In ref.~\cite{Kozaczuk:2012sx},
the signal expected from a large number of bubble collisions has been studied,
and it is found that the CMB data disfavor a bubble signal due to the low CMB quadrupole
power. We will also study this case, finding similar results.

In this paper we will focus on the technical details of our
analysis, presenting the main results in a companion paper~\cite{short}.
We restrict ourselves to the WMAP temperature data, and use
the WMAP7+BAO+$H_0$ cosmological parameters throughout~\cite{Komatsu:2010fb}.
In~\S\ref{sec:bubble_signal} we describe the calculation of the bubble signal,
in~\S\ref{sec:data_analysis} we describe our method, including the calculation of the likelihood, and describe
the Bayesian and Frequentist approaches, in \S\ref{sec:lots_of_bubbles} we extend the analysis to include
a large number of bubbles, and we conclude in \S\ref{sec:conc}.


\section{Mini-summary of the theory of bubble collisions}

In this section we briefly review the theory of bubble collisions,
focusing on aspects which will be needed for the data analysis.
For a detailed review, see~\cite{Kleban:2011yc}.

\begin{figure}[t]
  \begin{minipage}[b]{0.45\linewidth}
    \centering
    \includegraphics[width=\textwidth, clip=true, trim=0.8cm 1.4cm 0.9cm 2cm]{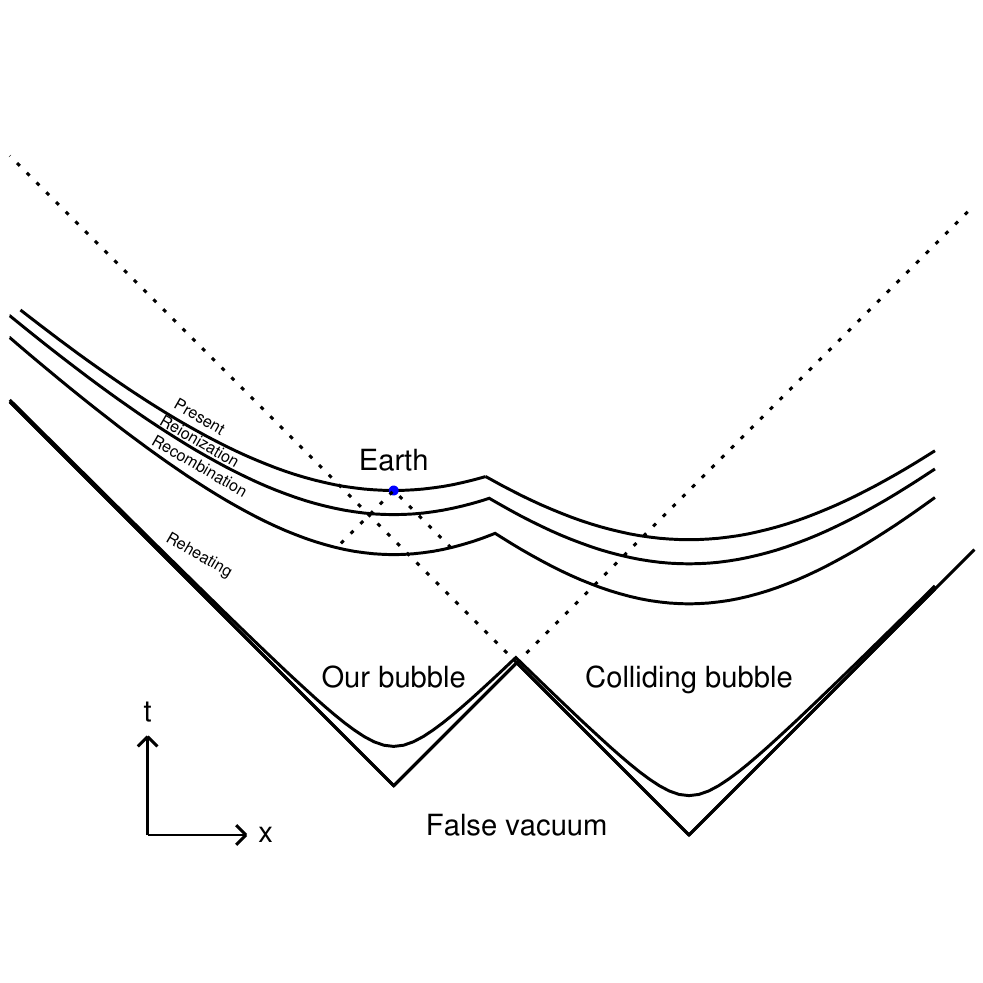}
    \caption{\label{fig:space-time} A spacetime diagram showing the causal structure of a
	single bubble collision, based on Fig. 3 from~\cite{Kleban:2011pg}.
	Coordinates are chosen so that light propagating in the plane of the diagram moves
	along $45^{\circ}$ lines.}
  \end{minipage}
  \hspace{0.5cm}
  \begin{minipage}[b]{0.45\linewidth}
    \centering
    \includegraphics[width=\textwidth, clip=true, trim=0.4cm 0.76cm 0.2cm 1.2cm]{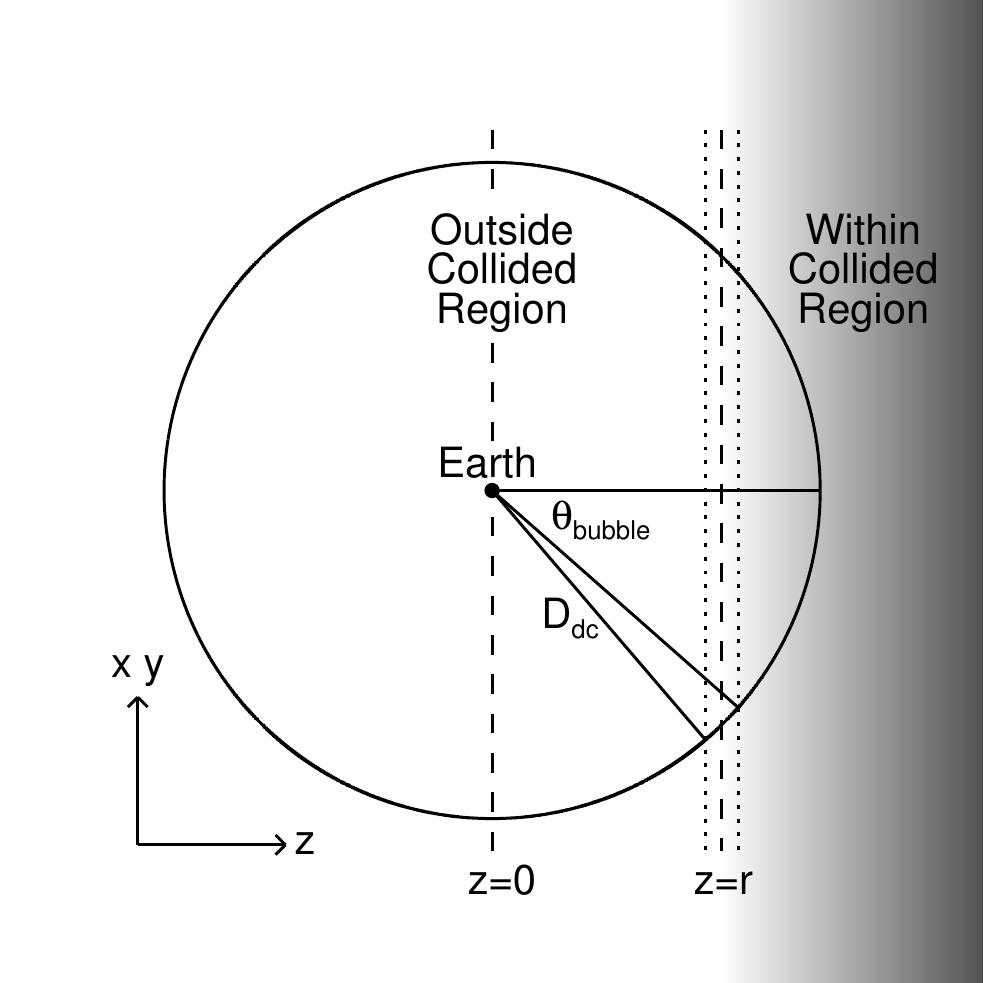}
    \caption{\label{fig:lss} The Earth's last scattering surface at the time of decoupling,
	based on Fig. 4 from~\cite{Kleban:2011pg}.
	The color of the shaded region indicates the magnitude of the curvature perturbation,
        assumed to be of the ``ramp'' form given below in Eq.~(\ref{eq:ramp_def}).}
  \end{minipage}
\end{figure}

The spacetime diagram for a single bubble collision is shown in Fig.~\ref{fig:space-time}.
When a bubble nucleates, the region of spacetime contained in the future lightcone of
the nucleation is an inflationary spacetime with small negative curvature.
Our observable Hubble volume is a smaller region given by taking the intersection with
the past lightcone of a present-day observer.
If there is a second bubble nucleation, and its future lightcone intersects our observable
Hubble volume, then the metric will be perturbed, and this will generate a disc-shaped CMB 
temperature perturbation~\cite{Aguirre:2007an,Chang:2008gj,Czech:2010rg,Kleban:2011yc,Gobbetti:2012yq}.

The expected number $N$ of bubble collisions is given by~\cite{Freivogel:2009it}
\be
\langle N\rangle\simeq \gamma \sqrt{\Omega_k} \frac{H_f^2}{H_i^2}\ .
\ee
Here $\sqrt{\Omega_k}$ represents the curvature of the current universe. It scales as $e^{-(N_i-N_\star)}$,
with $N_i$ the number of $e$-foldings during inflation and $N_\star\sim 63$ the number of $e$-foldings since
reheating. Unless there is a mechanism forcing the universe to have the lowest possible number of
inflationary $e$-foldings, then
$\sqrt{\Omega_k}$ is naturally an exponentially small number.
$\gamma$ is related to the decay rate per unit four-volume, $\Gamma$,
as $\gamma\sim \Gamma/H_f^4\sim e^{-S_E}$, where $H_f$ is the Hubble rate in the false vacuum region,
and $S_E$ is the Euclidan action of the instanton that mediates the decay of the false vacuum. $\gamma$
is expected to be an exponentially small number. Finally, $H_i$ is the inflationary Hubble rate during the
period of slow-roll inflation within our bubble. The ratio $H_f/H_i$ can be very large, if the vacuum
energy of the false vacuum is much larger than the one that drives slow roll inflation in our bubble. Of
course $H_i$ could be close to $H_f$, in which case there is no large factor. Summarizing, we have the product of
two exponentially small numbers and one possibly large number. The zeroth order expectation is that
$\langle N\rangle$ is exponentially small. A first order, more hopeful, expectation uses the possible
enhancement of $H_f/H_i$ to conclude that $\langle N\rangle$ is either very small or very large.  It therefore appears unlikely to us that $\langle N \rangle$ should be of order 1.  However, the theoretical understanding of eternal inflation and of the string landscape is still in a very preliminary stage. In particular it is possible that in the landspace the probability is dominated by decays with a small Euclidean action and by  local inflationary patches with small number of $e$-foldings. It is therefore hard to make strong statements about the theoretical expectations on $\langle N\rangle$. It is fair to say that the signal we are looking for is well defined, and since a detection would have such important theoretical implications, we proceed anyway.

Similar considerations apply for the expected typical size of the temperature perturbation induced by
the bubble collision. The bubble collision can be thought of as inducing a discontinuity $(\Delta\phi)_{\rm in}$
in the initial conditions of the inflaton $\phi$ in our patch. This initial discontinuity evolves during inflation
into what we later call the ramp profile, with typical temperature perturbation
$\zeta\sim (\Delta\phi)_{\rm in} \, (H_0/H_i) \, a_0 x_c e^{-(N_i-N_\star)}$, where $a_0$ and $H_0$ are the
current values of the scale factor and Hubble rate, respectively, and $x_c$  is the comoving
distance from the bubble wall to the interior~\cite{Gobbetti:2012yq}. As was the case for the expected number
of bubble collisions, the size of the perturbation can be very different from $10^{-5}$, which is the range
we will probe with a dedicated analysis. Conditioning on the fact that inflation needs to have happened can
motivate excluding very large density perturbations, but it does not exclude very small ones. Still, as for
the case of the number of collisions, the theoretical understanding of the probability for the amplitude of the signal is very preliminary, and the discovery of an event would have such importance that we
implement a dedicated analysis. Note too that the analysis we perform involves the development of techniques
that have a wide range of alternative applications.

Throughout this paper, we will parametrize the bubble size either by the comoving distance $r$ to the bubble
wall, or by the angular radius $\thetabubble$, which we define by:
\be
r=D_{dc}\cos{\thetabubble}
\ee
where $D_{dc}$ is the comoving distance to last scattering.  A bubble will be parametrized by its size (either $r$ or $\thetabubble$), its
angular location $\n$ on the sky (a unit two-vector), and one or more amplitude parameters (to be defined in~\S\ref{ssec:ramp_step}).

Although the number density of bubbles and their amplitudes depend on microphysics of the inflationary
model, the random distribution of bubble locations and sizes is determined by symmetry alone~\cite{Freivogel:2009it}.
The angular location $\n$ of the bubble is uniformly distributed over the sky, and in
the spatially flat limit $\Omega_k \ll 1$ (which we will assume throughout this paper),
the size parameter $r$ is a uniformly distributed random variable.
Equivalently, the size parameter $\thetabubble$ has the distribution:
\be
dP \propto dr \propto \sin(\thetabubble) d\thetabubble   \label{eq:size_prior}
\ee
This distribution implies that most of the bubbles are expected to have a large angular size.


\section{Bubble signal}
\label{sec:bubble_signal}

\subsection{Ramp and step models}
\label{ssec:ramp_step}

The bubble collision generates a contribution to the initial adiabatic curvature perturbation $\zeta$,
which evolves to generate a contribution to the CMB temperature anisotropy $a_{\ell m}$.
The spacetime symmetries of the collision imply that $\zeta$ is invariant under a residual SO(2,1) symmetry.
For a suitable choice of coordinates, and in the limit of zero spatial curvature,
the symmetry generators can be taken to be rotation around the $z$-axis and translations in the $x$ and $y$ directions.
Thus $\zeta(x,y,z)$ must be a function of $z$ alone.

In the simplest scenario, the bubble contribution to $\zeta$ is of the ``ramp'' form:
\be
\zeta(x,y,z) = \left\{ \begin{array}{cl}
  a^{\rm ramp} (z-r) & \mbox{if $z \ge r$}  \\
    0 & \mbox{if $z < r$}
\end{array} \right.   \label{eq:ramp_def}
\ee
where $r$ is the comoving distance to the bubble wall and $a^{\rm ramp}$ is a free parameter with units Mpc$^{-1}$.
The distance $r$ is related to the angular size $\thetabubble$ of the bubble by $\cos{\thetabubble} = r/D_{\rm dc}$.
We will refer to model~(\ref{eq:ramp_def}) as the ``ramp model''.
                                   
In addition to the ramp perturbation, there is a contribution from the signal at the boundary of the collision.
Following~\cite{Feeney:2010dd,Feeney:2010jj}, we will allow for a simple step function:
\be
\zeta(x,y,z) = \left\{ \begin{array}{cl}
  a^{\rm ramp} (z-r) + a^{\rm step} & \mbox{if $z \ge r$}  \\
    0 & \mbox{if $z < r$}
\end{array} \right.   \label{eq:step_def}
\ee
where $a^{\rm step}$ is dimensionless.
We will refer to model~(\ref{eq:step_def}) as the ``ramp+step model''.

The size of the boundary and the signal expected in the boundary region are not well understood theoretically,
and there is some disagreement in the literature as to which of (\ref{eq:ramp_def}) or (\ref{eq:step_def})
is better motivated.
In this paper, our focus is data analysis and we will not weigh in on this theoretical issue; we will simply show how to
perform the optimal analysis for both the ramp and ramp+step models.

\subsection{CMB temperature profiles}
\label{ssec:profiles}

To calculate the angular CMB temperature profile produced by the bubble, we
must account for the acoustic and gravitational physics which generates the CMB temperature anisotropy
from the initial curvature $\zeta$.
We can define a transfer function $\Delta_\ell(k)$ which represents the 
contribution of a Fourier mode of $\zeta$ with wavenumber $k$ to the
CMB temperature anisotropy at angular wavenumber $\ell$~\cite{Hu:1994jd,Hu:1994uz,Creminelli:2005hu}.
The CMB temperature, $a_{\ell m}$, is related to the 3D curvature perturbation $\zeta$ by
\be
\label{eq:almfromzeta}
a_{\ell m} = 4\pi i^{\ell} \int \frac{d^3\k}{(2\pi)^3} \, \tzeta(\k) \, \Delta_\ell(k) \, Y_{\ell m}^*(\hk)
\ee
where $\tzeta(\k) = \int d^3\x\, \zeta(\x) e^{-i\k\cdot\x}$ is the 3D Fourier transform
of $\zeta$.
In the special case where $\zeta$ is a function of $z$ alone, we have
$\tzeta(\k) = \tzeta(k_z) (2\pi)^2 \delta(k_x) \delta(k_y)$, where $\tzeta(k_z) = \int dz\, \zeta(z) e^{-ik_zz}$
is the 1D Fourier transform of $\zeta(z)$. Plugging into Eq.~(\ref{eq:almfromzeta}), we get:
\be
a_{\ell m} = b_\ell \, \sqrt{\frac{2\ell+1}{4\pi}} \delta_{m0}  \label{eq:alm_bubble_z}
\ee
where we have defined
\be
b_\ell = 2 \int_{0}^\infty dk_z\, \Delta_\ell(k_z) \left[ i^\ell \tzeta(k_z) + (-i)^\ell \tzeta(k_z)^* \right]  \label{eq:transfer_integral}
\ee
If the bubble collision is in direction $\n$ (rather than in the $z$-direction), then Eq.~(\ref{eq:alm_bubble_z})
for the temperature profile generalizes to:
\be
a_{\ell m} = b_\ell \, Y_{\ell m}^*(\n)   \label{eq:alm_bubble_general}
\ee
This can be shown by starting with Eq.~(\ref{eq:alm_bubble_z}) for a bubble in the $z$-direction,
and applying a rotation $R(\n)$ which carries direction $\hat z$ to direction $\n$:
\be
\label{eq:wigner_rotation}
[R(\n) \; a]_{lm} = \sum_{m^{\prime}} D^l_{mm^{\prime}} (\n) \; a_{lm^{\prime}}
\ee
where $D^l_{mm^{\prime}}$ is the Wigner D-matrix (for a definition see, for example,~\cite{McEwen:2006ke}).
Using the identity $D^l_{m0}(\n) = \sqrt{4\pi/(2\ell+1)} Y_{\ell m}^*(\n)$, we obtain Eq.~(\ref{eq:alm_bubble_general}) above.

As an aside, we note that the general form~(\ref{eq:alm_bubble_general}) is valid for any azimuthally symmetric profile,
even one which does not come from an initial curvature perturbation (e.g.~SZ emission from a galaxy cluster).
In general, the harmonic-space profile $b_\ell$ and its real-space counterpart $b(\theta)$ are related by:
\begin{eqnarray}
b(\theta) &=& \sum_\ell b_\ell \left( \frac{2\ell+1}{4\pi} \right) P_\ell(\cos\theta) \nn \\
b_\ell &=& 2\pi \int d(\cos\theta) P_\ell(\cos\theta) b(\theta)  \label{eq:bl_transform}
\end{eqnarray}
For the ramp and step bubble models, we can specialize the general expression~(\ref{eq:transfer_integral})
to obtain explicit formulas for $b_\ell$.
The relevant 1D Fourier transforms are $\tzeta_{\rm ramp}(k_z) = -e^{-ik_zr}/k_z^2$ and
$\tzeta_{\rm step}(k_z) = -ie^{-ik_zr}/k_z$.  We get:
\begin{eqnarray}
b_\ell^{\rm ramp}(r) &=& \left\{
\begin{array}{cl}
4 \, (-1)^{\ell/2+1}   \int_0^\infty dk\, k^{-2} \Delta_\ell(k) \, \cos(kr) & \mbox{for even $\ell$} \\
4 \, (-1)^{(\ell+1)/2} \int_0^\infty dk\, k^{-2} \Delta_\ell(k) \, \sin(kr) & \mbox{for odd  $\ell$}
\end{array}
\right. \\
b_\ell^{\rm step}(r) &=& \left\{
\begin{array}{cl}
4 \, (-1)^{\ell/2+1} \int_0^{\infty} dk\, k^{-1} \Delta_\ell(k) \, \sin(kr)  & \mbox{for even $\ell$} \\
4 \, (-1)^{(\ell-1)/2} \int_0^\infty dk\, k^{-1} \Delta_\ell(k) \, \cos(kr)  & \mbox{for odd  $\ell$}
\end{array}
\right.
\end{eqnarray}
We compute these profiles numerically, using CAMB~\cite{Lewis:1999bs} to compute the transfer function $\Delta_\ell(k)$.

Note that our normalization convention is to parametrize the bubble amplitude by parameters $a^{\rm ramp}$, $a^{\rm step}$
which appear in Eqs.~(\ref{eq:ramp_def}),~(\ref{eq:step_def}) for the curvature perturbation $\zeta$,
and have units [Mpc$^{-1}$] and [dimensionless] respectively (no $\mu$K).
In particular this means that a bubble with positive amplitude corresponds to a cold spot 
on the sky (provided the radius is $\gsim 1^\circ$), since positive $\zeta$ corresponds to negative $\Delta T$ on large scales.
We elaborate on the relation between $\zeta$ and $\Delta T$ in the next section.

\subsection{Can CMB transfer functions be neglected?}
\label{ssec:cmb_transfer}

The expressions from the previous section for the profiles $b_\ell^{\rm ramp}$, $b_\ell^{\rm step}$ are
nontrivial to evaluate, and one may wonder whether it is a good approximation to simply assume that the
temperature perturbation $\Delta T$ is proportional to the value of $\zeta$ on the last scattering surface (the
Sachs-Wolfe approximation).

\begin{figure}[t]
  \begin{center}
  \includegraphics[width=10cm, clip=true, trim=0.5cm 0.4cm 0.5cm 0.35cm]{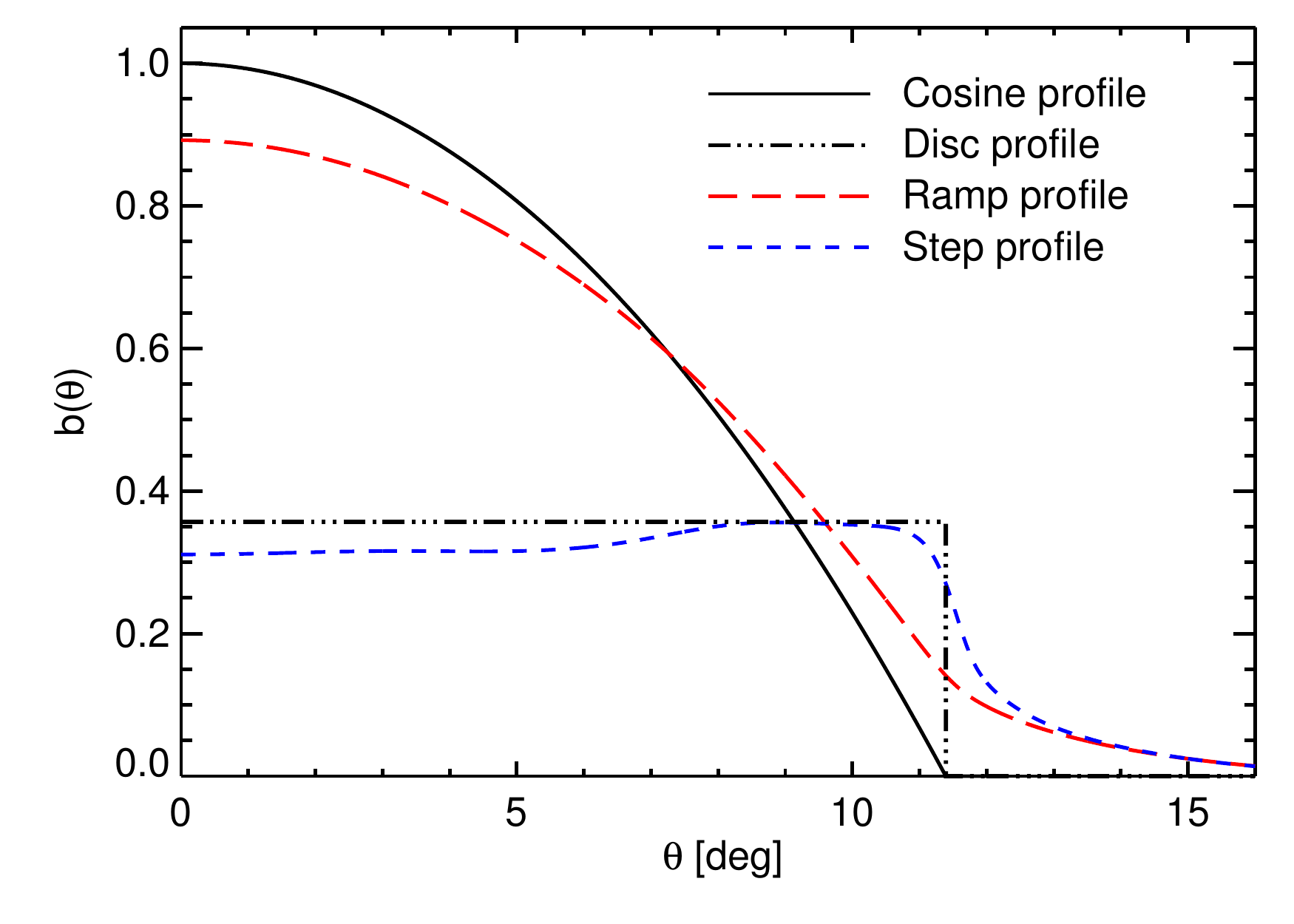}
  \caption{\label{fig:bubble_profile} Angular bubble profiles $b(\theta)$ defined in~\S\ref{ssec:profiles},
for bubbles at comoving distance $r=13886.6\,$Mpc (corresponding to angular size
$\theta_{\rm bubble} = 11.39^{\circ}$) and arbitrary normalization.
The ramp and step profiles are obtained by evolving an initial feature in the adiabatic
curvature perturbation (Eqs.~(\ref{eq:ramp_def}),~(\ref{eq:step_def})) forward to obtain a CMB temperature
profile, using the CMB transfer function.
The cosine and disc profiles are approximations to the ramp and step profiles in which the CMB transfer function
is omitted (Eq.~(\ref{eq:cosine_disc_def})).
Qualitatively, the effect of including the CMB transfer function is to smooth sharp features
in the profile, with smoothing length $\approx 1^\circ$ given by the CMB acoustic scale.}
  \end{center}
\end{figure}

To study this question quantitatively, we define ``cosine'' and ``disc'' profiles in real space by:
\begin{eqnarray}
b_{\rm cosine}(\theta) &=& \left\{ \begin{array}{cl}
     \cos{\theta}-\cos{\thetabubble}  & \mbox{if $\theta \le \thetabubble$} \\
      0  & \mbox{if $\theta \ge \thetabubble$}
\end{array} \right. \nn \\
b_{\rm disc}(\theta) &=& \left\{ \begin{array}{cl}
      1  & \mbox{if $\theta \le \thetabubble$} \\
      0  & \mbox{if $\theta \ge \thetabubble$}
\end{array} \right. \label{eq:cosine_disc_def}
\end{eqnarray}
where $\thetabubble$ is the angular size of the bubble.
We note that the integral~(\ref{eq:bl_transform}) can be evaluated analytically for these profiles,
giving the following harmonic-space profiles:
\begin{eqnarray}
b_\ell^{\rm cosine} &=& 2\pi \left(
    \frac{P_{\ell+2}(z)}{(2\ell+1)(2\ell+3)} 
  - 2 \frac{P_\ell(z)}{(2\ell-1)(2\ell+3)}
  + \frac{P_{\ell-2}(z)}{(2\ell-1)(2\ell+1)}
\right)  \nn \\
b_\ell^{\rm disc} &=& 2\pi \left( \frac{P_{\ell-1}(z) - P_{\ell+1}(z)}{2\ell+1} \right)
\end{eqnarray}
where $z=\cos{\thetabubble}$.
In Fig.~\ref{fig:bubble_profile}, we show the ramp, step, cosine, and disc profiles
for a bubble at a distance $r = 13886.6\,$Mpc, corresponding to angular size
$\theta_{\rm bubble} = 11.39^{\circ}$, with arbitrary normalizations.

In the Sachs-Wolfe approximation, the CMB temperature in direction $\n$ is given by
$\Delta T(\n) = -\Phi(D_{\rm dc} \n)/3$, where $\Phi$ is the Newtonian
potential.
On large scales, $\Phi$ is related to $\zeta$
by $\Phi = (3+3w)/(5+3w) \zeta$, where $w\approx 0.11$ is the equation of state
parameter at last scattering.  Putting this together, the Sachs-Wolfe approximation
applied to the bubble profile reads:
\bea
b_\ell^{\rm ramp} & \approx & -\frac{1}{3} \left( \frac{3+3w}{5+3w} \right) D_{\rm dc} b_\ell^{\rm cosine}  \label{eq:sw_ramp} \\
b_\ell^{\rm step} & \approx & -\frac{1}{3} \left( \frac{3+3w}{5+3w} \right) b_\ell^{\rm disc}  \label{eq:sw_step}
\eea
Considering the case of the ramp profile first, we find that the Sache-Wolfe approximation~(\ref{eq:sw_ramp}) is
excellent for most bubble sizes, but breaks down for bubbles which are very small or very large.
For very small bubbles (i.e.~$\thetabubble \lsim 1^\circ$ or equivalently $r\approx D_{\rm dc}$), the bubble profile is widened by $\approx 1^\circ$
due to acoustic physics encoded in the CMB transfer function.
For very large bubbles (i.e.~$\thetabubble \approx 90^\circ$ or $r \ll D_{\rm dc}$),
the Sachs-Wolfe approximation breaks down due to the contribution from the ISW effect, which
turns out to partially cancel the Sachs-Wolfe contribution and reduces the bubble amplitude
by $\approx 30$\%.

For the step profile, we find that the Sachs-Wolfe approximation~(\ref{eq:sw_step}) is not very accurate.
More quantitatively, the correlation coefficient $r(b_\ell^{\rm step},b_\ell^{\rm disc})$ between the step and disc profiles
is never close to 1; we find $0.2 \le r \le 0.7$ depending on the bubble radius.
We give the precise definition of $r(b,b')$ in Eq.~(\ref{eqn:rbb}) below,
but for now we simply treat it as a metric that can take values between -1 and 1,
with correlation $\pm 1$ meaning that the profiles are identical up to rescaling.

When we search for the ramp or step profiles in CMB maps,
the ramp profile gets most of its statistical weight from low $\ell$, whereas the step profile
gets its statistical weight from the full range of $\ell$ values which are measured with
appreciable signal-to-noise.
Intuitively, when we search for the profiles against the CMB sky,
most of the signal for the ramp profile comes from the ``bulk'' of the profile,
whereas most of the signal for the step profile comes from the sharp drop near the edge.

Summarizing, CMB transfer functions are an order-one effect for the step profile and
must be included, but are less important for the ramp profile.
This makes intuitive sense because the ramp profile is mainly a low-$\ell$ signal, the
step profile is mainly a high-$\ell$ signal, and transfer functions are unimportant
at low $\ell$.
However, even for the ramp profile, transfer functions can affect the shape or amplitude
of bubbles with either $\thetabubble\approx 0$ or $\thetabubble\approx 90^\circ$, and should
be included in a precise analysis.\footnote{We 
also note that for an experiment which measures the damping
tail $\ell\gsim 2000$, the Sachs-Wolfe approximation for the ramp profile
cannot be used without modification, since it would give $b_\ell$'s which are 
not exponentially suppressed at high $\ell$, which leads to a spurious
detectable high-$\ell$ signal since the CMB power spectrum $C_\ell$ is exponentially suppressed.}

\subsection{Discretizing the bubble radius}
\label{ssec:discretizing}

The data analysis algorithms in the next section will require the bubble radius, or
equivalently the distance $r$ to the bubble wall, to be discretized to a 
finite set of values $r_1, \cdots, r_N$.
If we use too few $r$ values then we will decrease our statistical power, since we
will end up searching for an incomplete set of profiles in the data.
If we use too many $r$ values then the computational cost increases (the cost will
turn out to be roughly proportional to $N$).
We therefore require a procedure to determine the minimum number of $r$ values which
are needed.

To motivate our choice for this procedure, we first consider the (unnormalized) minimum
variance estimator for the amplitude of a bubble at known location $\n_0$:
\be
\E = \sum_{\ell m} \frac{b_\ell}{C_\ell + N_\ell} a_{\ell m} Y_{\ell m}(\n_0)
\ee
where we have assumed all-sky homogeneous noise with power spectrum $N_\ell$.
The covariance $\mbox{Cov}(\E,\E')$ of two such estimators with profiles $b_\ell$ and $b'_\ell$ is
given by:
\be
b \cdot b' = \sum_\ell (2\ell+1) \frac{b_\ell b'_\ell}{C_\ell+N_\ell}  \label{eqn:dot_product}
\ee
where we have introduced a dot product notation for compactness.

We determine a minimal set of $r$ values by requiring that adjacent $r$ values be ``close'',
in the sense that the associated profiles have a correlation coefficient greater than 0.97,
where the correlation is defined using the dot product in Eq.~(\ref{eqn:dot_product}):
\be
\label{eqn:rbb}
r (b,b') = \frac{b\cdot b'}{\sqrt{(b\cdot b) (b'\cdot b')}} > 0.97
\ee
This definition of closeness corresponds to observational indistinguishability: two profiles
are close if they cannot be distinguished statistically given a noisy observation of the CMB.
Note that we are implicitly approximating the real WMAP noise model by all-sky homogeneous
noise with the same noise power spectrum, but this is a reasonable approximation if we just
want to decide whether two profiles are highly correlated.

For the ramp profile we find we need 129 $r$ values to cover the range $r_{\rm min}=0$
to $r_{\rm max}=14165.64$~Mpc.
For the step profile, we find that 2002 $r$ values are needed, due to smaller-scale features in
the profile.

\section{Data analysis}
\label{sec:data_analysis}

In this section, we consider the question: what is the optimal statistic for detecting a
bubble collision, and how can we evaluate it in a computationally feasible way?
We will answer this question in both the Bayesian (\S\ref{ssec:bayesian}) and frequentist 
(\S\ref{ssec:frequentist}) statistical frameworks.
In both cases, we will find that there is a natural choice of optimal statistic, but computing it
appears to be prohibitively expensive.  However, we will find computational tricks which will allow
us to evaluate the optimal statistic with reasonable computational cost, and without making any
approximations.

\subsection{Definitions and notation}
\label{ssec:notation}

We denote the noisy CMB data by a vector $d_{\mu p}$, where the index $\mu$ runs over observing channels (in our WMAP
pipeline, $\mu$ indexes one of the six differential assemblies V1,V2,W1,W2,W3,W4) and $p$ runs over sky pixels.
We keep the observing channels
separate, rather than combining them to a single map $d_p$, to avoid making the analysis suboptimal by
degrading all channels to the resolution of the worst channel.  Throughout this paper, we refer to a two-index
object such as $d_{\mu p}$ as a ``per-channel pixel-space map'', and a single-index object $x_p$ as a ``pixel-space map''.

Let $c_{\ell m}$ denote the CMB realization, and let $A_{\mu p,\ell m}$ be the operator which acts
on the harmonic-space map $c_{\ell m}$ to produce a per-channel pixel-space map $(Ac)_{\mu p}$.
In detail, $(Ac)$ is defined by multiplying $c$ by the beam transfer function for each
channel (and the Healpix window function), then applying spherical harmonic transforms
to obtain a per-channel pixel-space map.

We define a covariance matrix $C = S+N$ for the data vector $d_{\mu p}$, consisting of signal and noise contributions.
The covariance matrix is defined in the per-channel pixel-space domain; thus it has indices $C_{\mu p,\mu'p'}$.
In WMAP, the noise in different pixels is uncorrelated to an excellent approximation, and so we treat
the noise covariance as diagonal: $N_{\mu p, \mu'p'} = \sigma_{\mu p}^2 \delta_{\mu\mu'} \delta_{pp'}$,
where $\sigma_{\mu p}^2$ is a noise variance which can depend on channel $\mu$ and sky pixel $p$.
The signal covariance is given by $S = A \Sigma A^T$, where the matrix $\Sigma$ is diagonal in harmonic space:
$\Sigma_{\ell m, \ell'm'} = C_\ell \delta_{\ell\ell'} \delta_{mm'}$.
Note that with this covariance matrix $S$, the signal in different channels $\mu,\mu'$ is 
100\% correlated (but convolved with different beams).

We will shortly encounter expressions involving the inverse covariance matrix $C^{-1}$.
The matrix size is much too large to invert (or even store) $C$ in dense form.
However, given a per-channel pixel-space map $x$, there are iterative algorithms which can 
efficiently compute the matrix-vector product $C^{-1}x$.
We will use the multigrid conjugate gradient algorithm from~\cite{Smith:2007rg}, which
can perform one $C^{-1}$ multiplication in approximately 15 core-minutes for WMAP.
Some technical details of the $C^{-1}$ filter are presented in Appendix~\ref{app:c_inverse}.
For now, we just remark on one important feature: we can include a pixel mask
by assigning formally infinite noise variance to masked pixels, and similarly 
include monopole+dipole marginalization
by suitably modifying the noise covariance so that
the relevant modes are given infinite variance.

In order to present our algorithms in maximum generality, we introduce a general
notation for azimuthally symmetric profiles as follows.  
We assume that the profile is parametrized by a direction $\n$, a vector of ``linear''
parameters $a_1, \cdots, a_M$, and a discrete index $I = 1, 2, \cdots, N$
representing one or more ``nonlinear'' parameters which have been discretized.
For the bubble collision problem, the linear parameters $a_i$ represent profile amplitudes,
i.e.~either $a^{\rm ramp}$ or $a^{\rm step}$.
The number of linear parameters $M$ is equal to 1 for the ramp model and equal to 2 for the ramp+step model.
The index $I$ represents the bubble radius, discretized as described in \S\ref{ssec:discretizing}.
We will construct optimal statistics to search for profiles of the parametrized form:
\be
a_{\ell m} = \sum_{i=1}^M a_i b_\ell^{Ii} Y_{\ell m}^*(\n)   \label{eq:general_profile}
\ee
This parametrization should apply generally to any azimuthally symmetric family of profiles,
for example textures~\cite{Feeney:2012jf} or SZ clusters.
In the subsequent sections, we will indicate which parts of our analysis framework apply in this generality,
and which parts are specific to the case of the bubble collision.

As a final piece of notation, we define the harmonic-space map $\beta_{Ii\n}$ by:
\be
(\beta_{Ii\n})_{\ell m} = b^{Ii}_\ell Y_{\ell m}^*(\n)
\ee
so that the harmonic-space profile in Eq.~(\ref{eq:general_profile}) is equal to $\sum_i a_i \beta_{Ii\n}$.
Less formally, $\beta_{Ii\n}$ is the profile of a bubble with location $\n$, size $I$, amplitude parameter
$a_i=1$, and amplitude parameters $a_j=0$ for $j\ne i$.

\subsection{An algorithm for fast calculation of $\Delta\chi^2$}
\label{ssec:chi2}

Given a data realization $d_{\mu p}$, we define its $\chi^2$ by $\chi^2(d) = d^T C^{-1} d$.
Note that the likelihood function for the signal+noise realization $d$ is a multivariate Gaussian
with covariance matrix $C$:
\be
\L(d) = \mbox{det}(2\pi C)^{-1/2} \exp\left( -\frac{1}{2} d^T C^{-1} d \right)  \label{eq:gaussian_likelihood}
\ee
so the $\chi^2$ and the likelihood are related by $\L(d) \propto \exp(-\chi^2(d)/2)$.

Given data realization $d_{\mu p}$ and profile parameters $(\{a_i\},I,\n)$ we define
$\Delta\chi^2(d,\{a_i\},I,\n)$ to be the change
in $\chi^2$ when the profile is subtracted from the data realization $d$:
\be
\Delta\chi^2(d,\{a_i\},I,\n) = \chi^2\!\left( d - \sum_{i=1}^M a_i A \beta_{Ii\n} \right) - \chi^2(d)   \label{eq:delta_chi2_def}
\ee
The Bayesian and frequentist statistics in the next two sections will require very fast evaluation
of $\Delta\chi^2(d,\{a_i\},I,\n)$, so in this section we will give an algorithm for computing it.
The algorithm is organized as a set of precomputations which subsequently allow $\Delta\chi^2$ to
be evaluated at a single point $(d,\{a_i\},I,\n)$ via an $\bigoh(1)$ table lookup.

We begin by rearranging (\ref{eq:delta_chi2_def}) to write $\Delta\chi^2$ in the following form:
\be
\Delta\chi^2(d,\{a_i\},I,\n) = -2 \sum_i a_i \, \beta_{Ii\n}^T A^T C^{-1} d 
   + \sum_{ij} a_i a_j \beta_{Ii\n}^T A^T C^{-1} A \beta_{Ij\n}   \label{eq:delta_chi2_2term}
\ee
Let us consider the two terms separately.
The first term in Eq.~(\ref{eq:delta_chi2_2term}), the ``data-bubble'' term, can be computed efficiently as follows.
Define pixel-space maps $D_{Ii}(\n)$ by:
\be
D_{Ii}(\n) = \sum_{\ell m} b^{Ii}_\ell (A^T C^{-1} d)_{\ell m} Y_{\ell m}^*(\n)  \label{eq:DIi_def}
\ee
We precompute the maps $D_{Ii}(\n)$ and save them to disk, by first
computing the harmonic-space map $A^T C^{-1} d$ using the fast $C^{-1}$ multiplication algorithm,
and then calculating the pixel-space map $D_{Ii}(\n)$ directly from Eq.~(\ref{eq:DIi_def})
using a spherical harmonic transform for each pair $(I,i)$.
The data-bubble term in Eq.~(\ref{eq:DIi_def}) is equal to $-2 \sum_i a_i D_{Ii}(\n)$, so after the $D_{Ii}(\n)$
maps have been precomputed, the data-bubble term can subsequently be evaluated with $\bigoh(1)$ computational cost.

Note that in this algorithm, we discretize the bubble location parameter $\n$ using a finite pixelization of the sphere.
In principle this pixelization need not have the same resolution as the pixelization used to represent
the CMB maps (although in our WMAP pipeline, we use an $\Nside=512$ Healpix pixelization throughout).

The second term in Eq.~(\ref{eq:delta_chi2_2term}), the bubble-bubble term, cannot be calculated using the same trick
since this would require computing $C^{-1} A \beta_{Ii\n}$ for every pair $(I,i)$ and pixel $\n$.  However, the
bubble-bubble term can be calculated statistically.
Suppose that we simulate a per-channel map $x_{\mu p}$ with the same statistical properties as the data, so that
the covariance matrix $\langle x x^T \rangle$ is equal to $C$.
Let $X_{Ii}(\n)$ be the pixel-space map which is obtained from the simulation $x$ in the same way
that the map $D_{Ii}(\n)$ above is obtained from the data:
\be
X_{Ii}(\n) = \sum_{\ell m} b^{Ii}_\ell (A^T C^{-1} x)_{\ell m} Y_{\ell m}^*(\n)
\ee
Then the Monte Carlo average $\langle X_{Ii}(\n) X_{Ij}(\n) \rangle$, taken over random realizations of $x$, is given by:
\be
\langle X_{Ii}(\n) X_{Ij}(\n) \rangle = \beta_{Ii\n}^T A^T C^{-1} A \beta_{Ij\n}
\ee
The right-hand side is precisely what is needed to compute the bubble-bubble term in Eq.~(\ref{eq:delta_chi2_2term}).
Therefore, we use the following algorithm to compute the bubble-bubble term.
We do an outer loop over Monte Carlo simulations $x$, and for each simulation we compute the maps $X_{Ii}(\n)$
using the same algorithm that was used to compute the maps $D_{Ii}(\n)$ above, 
and accumulate the simulation's contribution to the Monte Carlo average maps $\langle X_{Ii}(\n) X_{Ij}(\n) \rangle$.
After all Monte Carlos have been run, we write the maps $\langle X_{Ii}(\n) X_{Ij}(\n) \rangle$ to disk.
After precomputing these maps, the bubble-bubble term can be evaluated at a point $(d,\n,\{a_i\},I)$
with $\bigoh(1)$ computational cost.

This concludes our fast algorithm for calculating $\Delta\chi^2$.
The intuitive idea behind the algorithm is that we avoid doing a brute force scan over model parameters $(\n,\{a_i\},I)$
whenever possible.
An obvious optimization is to eliminate the scan over the linear parameters $\{a_i\}$ by noting that $(\Delta\chi^2)$
is a quadratic polynomial in these parameters, so it suffices to compute the coefficients of the polynomial
(i.e.~the maps $D_{Ii}(\n)$ and $\langle X_{Ii}(\n) X_{Ij}(\n) \rangle$).
What is less obvious is that the scan over the location parameter $\n$ can be eliminated.  In our algorithm we effectively
evaluate all $\n$ simultaneously, using some algebraic tricks and a Monte Carlo approach to the bubble-bubble term.
We do need to scan the bubble radius parameter (or more generally, any nonlinear parameters
represented by the discrete index $I$) but we can organize the computation so that we do not pay the computational
cost of a $C^{-1}$ multiplication for each value of $I$, only the cost of a single spherical transform.

A final comment is that running our fast $(\Delta\chi^2)$ algorithm on a Monte Carlo ensemble of data realizations,
rather than a single realization $d$, is computationally feasible.
This is because the bubble-bubble term is the same in every Monte Carlo iteration, 
so we only need to recompute the data-bubble term.  Since the data-bubble term is much faster to compute than the
bubble-bubble term (by a factor equal to the number of Monte Carlos), evaluating $\Delta\chi^2$ on an 
ensemble of Monte Carlos has roughly the same computational cost as the precomputations which are needed
to evaluate it once.

\subsection{Bayesian analysis}
\label{ssec:bayesian}

In a Bayesian analysis framework, we start with a prior $p(\{a_i\},I,\n)$ on the parameters
of the model, and wish to compute the posterior likelihood $\L(\{a_i\},I,\n|d)$ given data 
realization $d$.
From the posterior, we can compute various derived quantities such as confidence regions 
and evidence integrals.

Using Bayes' theorem and the form of the Gaussian likelihood in Eq.~(\ref{eq:gaussian_likelihood}), 
the posterior likelihood is:
\begin{eqnarray}
\L(\{a_i\},I,\n|d) & \propto & \L(d|\{a_i\},I,n) \, p(\{a_i\},I,\n) \nn \\
  & \propto & \exp\left( -\frac{1}{2} \Delta\chi^2(d,\{a_i\},I,\n) \right) p(\{a_i\},I,\n)
\end{eqnarray}
Therefore, our fast $\Delta\chi^2$ algorithm from the preceding section lets us evaluate
the exact posterior as a table lookup with very minimal computational cost (tens of CPU cycles).
This makes the Bayesian analysis essentially trivial; for example confidence regions can be
determined by gridding the likelihood rather than using an MCMC.

The rest of this section is devoted to some practical details of the analysis for the special
case of a bubble collision (rather than an arbitrary azimuthally symmetric set of profiles).
In this case, the form of the prior $p(\{a_i\},I,\n)$ is
constrained by symmetry: all directions $\n$ are equally likely, and the comoving distance $r$
to the bubble wall is uniformly distributed (see Eq.~(\ref{eq:size_prior}) above).
Since the index $I$ represents distance to the bubble wall,
discretized to some set of values $r_1, r_2, \cdots, r_N$,
we take the prior on the discrete index $I$ to be proportional to
\be
\label{eqn:trapezoid}
w_I = \left\{
\begin{array}{cl}
(r_2-r_1)/2         & \mbox{for $I=1$} \\
(r_{I+1}-r_{I-1})/2 & \mbox{for $1 < I < N$} \\
(r_N - r_{N-1})/2     & \mbox{for $I=N$}
\end{array}
\right.
\ee
which corresponds to a uniform distribution in $r$, discretized with trapezoid rule weighting.
The prior $p(\{a_i\},I,\n)$ then factorizes:
\be
p(\{a_i\},I,\n) = \frac{w_I}{\Npix \, (r_N-r_1)} \, p(\{a_i\})  \label{eq:factorized_prior}
\ee
where $p(\{a_i\})$ is a prior on the amplitude parameters, which 
depends on detailed physics of the inflationary model and cannot be deduced from symmetry alone.
If a model-independent analysis is desired, there seems to be no particularly well-motivated choice
of prior $p(\{a_i\})$, and so we take a uniform prior for simplicity.

Given the form~(\ref{eq:factorized_prior}) for the prior, we can marginalize over the bubble location and
radius to obtain the posterior likelihood for the amplitude parameters:
\be
\L(\{a_i\}|d) \propto p(\{a_i\}) \sum_{I\n} \frac{w_I}{\Npix \, (r_N-r_1)} \exp\left( -\frac{1}{2} \Delta\chi^2(d,\{a_i\},I,\n) \right)  \label{eq:posterior}
\ee
In practical data analysis, this likelihood function needs a small modification for the following reason.
We must mask regions of high foreground emission such as the Galactic plane, and the above likelihood
sums over all bubble radii and locations, including bubbles which are completely masked and therefore
unconstrained by the data.  One symptom of this disease is that as the amplitude parameters $\{a_i\}$ are taken to
infinity, the likelihood is not exponentially suppressed, but approaches a nonzero constant, since
we can ``hide'' a bubble of arbitrary large bubble amplitude in the masked part of the sky.

For this reason, we omit pairs $(I,\n)$ in the sum~(\ref{eq:posterior}) for which the corresponding bubble is unconstrained.
If we simply excluded all pixels $\n$ which are in the Galactic mask, then we would be throwing away information.
If a bubble is centered on a masked pixel but a significant fraction of the bubble spills outside the mask, then
there are still many unmasked pixels that can be used to calculate a $\Delta\chi^2$ value. To account for this, we
exclude pairs $(I,\n)$ such that:
\be
\beta_{Ii\n}^T C^{-1} \beta_{Ii\n} < 0.1 \, \Big\langle \beta_{Ii\n'}^T C^{-1} \beta_{Ii\n'} \Big\rangle_{\n'}  \label{eq:center_mask}
\ee
where the mean $\langle\cdot\rangle_{\n'}$ is taken over high latitude pixels $\n'$ away from the galactic plane.
This radius-dependent criterion for masking bubble centers means that we only mask a bubble location $(I,\n)$
if $\approx 90$\% of the temperature profile is covered by the Galactic mask.

Omitting pairs $(I,\n)$ which correspond to masked bubbles simply means that our Bayesian likelihood
is the posterior likelihood for a bubble which is constrained to lie in the observable part of our sky, 
in the same sense that we constrain the bubble wall to intersect our observable Hubble volume.
Parameter constraints derived from this likelihood have a rigorous Bayesian interpretation as constraints on
bubbles which lie in our Hubble volume and are not obscured by the Galaxy, with bubbles outside this
observable volume unconstrained by the analysis.

We assign confidence regions using the posterior likelihood $\L(\{a_i\}|d)$ defined in Eq.~(\ref{eq:posterior}), 
and statistically test for bubbles by asking whether the point $\{a_i\}=0$ is contained in the 
95\% (for example) confidence region.
We define confidence regions corresponding to probability $p=0.95$ as follows.
If there is only one amplitude parameter $a$, we define the confidence region $[a_{\rm min}, a_{\rm max}]$
by the requirement that the total probability in each one-sided tail be $(1-p)/2$.  Formally,
\be
\int_0^{a_{\rm min}} \L(a|d) = \int_{a_{\rm max}}^\infty \L(a|d) = \frac{1-p}{2} \int_{-\infty}^\infty \L(a|d)  \label{eq:bayesian_confidence_regions1}
\ee
In two or more variables $\{a_i\}$, we threshold the likelihood
so that the total likelihood above threshold is $p$.
More formally, we define the confidence region by
solving for the value $\L_0$ such that the set of points $a_i$ satisfying $\L(a|d) \ge \L_0$
satisfies:
\be
\int_{\L(a|d) \ge \L_0} da_i \, \L(a|d) = p \int da_i \, \L(a|d)   \label{eq:bayesian_confidence_regions2}
\ee
and taking the confidence region to be the set of $a$-values such that $\L(a|d) \ge \L_0$.

Note that if we do not mask bubble centers, then the integrals in 
Eqs.~(\ref{eq:bayesian_confidence_regions1}),~(\ref{eq:bayesian_confidence_regions2})
diverge and we cannot define confidence regions.
Thus some prescription for bubble center masking seems to be necessary, although our prescription~(\ref{eq:center_mask})
is not the only possibility.

An alternate statistical test for bubbles is to compute the Bayesian evidence for the bubble model,
and compare it to the evidence for a no-bubble model with the usual Gaussian likelihood.
The relevant Bayes factor is:
\be
K = \frac{\int \{da_i\} \, p(\{a_i\}) \sum_{I\n} w_I \exp(-\frac{1}{2}\Delta\chi^2(d,\{a_i\},I,\n))}{\int \{da_i\} \, p(\{a_i\}) \sum_{I\n} w_I}
\ee
which is only defined if the prior $p(\{a_i\})$ is normalizable (i.e.~$\int \{da_i\} \, p(\{a_i\}) < \infty$).
For a given inflationary model, there should be a physically defined prior $p(\{a_i\})$ which is normalizable,
but there is no clear choice of normalizable prior which is model-independent.  For this reason, our perspective
is that the Bayesian evidence is not a good statistical test for bubbles in a generic, model-independent
analysis, and we use confidence regions instead.

The posterior likelihood $\L(\{a_i\}|d)$ in Eq.~(\ref{eq:posterior}) has the
following, perhaps counterintuitive, property which deserves explicit comment.
For a randomly generated realization, there is an order-one probability that the maximum likelihood is
very close to the point $\{a_i\}=0$, much closer than the width of the likelihood.
We discuss this further in Appendix~\ref{app:like_model} and develop an analytic model of the likelihood. 
We show that this behavior is expected, and such a realization should simply be interpreted
as one which is consistent with a Gaussian field, with no statistical evidence for bubbles.

\subsection{Frequentist analysis}
\label{ssec:frequentist}

In this section we will construct optimal frequentist statistics for the bubble collision problem,
and find an algorithmic trick which makes
the analysis computationally feasible.
We should say from the outset that we will
formulate the frequentist analysis as a procedure for defining confidence regions by hypothesis testing,
using a likelihood ratio test which can be shown to be optimal; we will not use the term ``estimator''.
In cosmology, frequentist analyses are often formulated in a different way, by
defining a global estimator for the model parameters (in this case the bubble amplitudes).
This type of estimator-based analysis is optimal in many cases (the formal criterion for optimality
is that the estimator should be a sufficient statistic for the likelihood ratio test), but fails
to capture key structure of the bubble collision problem.
To see this intuitively, consider a realization with multiple statistically significant bubble-like features.
The likelihood for the bubble amplitude now has multiple peaks and the correct confidence regions consist of multiple 
disconnected ``islands''.  This structure will not be captured if we try to compress the likelihood into
a single number by defining an estimator for the bubble amplitude.  Therefore, our focus will not be
on estimators, but on identifying the optimal Monte Carlo procedure for testing whether the data are
consistent with a given set of bubble amplitude parameters $\{a_i\}$.

First consider the question: how do we test whether the no-bubble model (i.e.~$\{a_i\}=0$)
is consistent with the data, at say 95\% confidence level?
In a frequentist analysis, confidence regions are defined by Monte Carlo based hypothesis testing.
We construct a test statistic $\rho_0$ which statistically separates no-bubble realizations from realizations
with bubbles, and evaluate $\rho_0$ on an ensemble of Monte Carlo simulations.
If the value of $\rho_0$ on the data is larger than 95\% of the simulations, then $\{a_i\}=0$ is excluded at
95\% CL.

Constructing optimal test statistics is sometimes a challenge in the frequentist approach, but
for the bubble collision problem there is a natural choice, as we now explain.
We will use the prior on the bubble location from the previous section:
all directions $\n$ are equally likely, and the distance to the bubble wall
is distributed with PDF proportional to the quantity $w_I$ defined in Eq.~(\ref{eqn:trapezoid}).
We will not need a prior $p(\{a_i\})$ on the amplitude parameters.

For a fixed set of amplitude parameters $\{a_i\}$, the conditional likelihood $\L(d|\{a_i\})$ is given by:
\be
\L(d|\{a_i\}) \propto \sum_{I\n} \frac{w_I}{\Npix \, (r_N-r_1)} \exp\left( -\frac{1}{2} \Delta\chi^2(d,\{a_i\},I,\n) \right)
\ee
Our frequentist test statistic $\rho_0$ will be a likelihood ratio statistic:
\be
\label{eqn:frequentist_test_statistic}
\rho_0(d) = \max_{\{a_i\}} \frac{\L(d|\{a_i\})}{\L(d|0)}
\ee
If $\rho_0(d)$ is larger than the value found in a large fraction (say 95\%) of the simulations,
then we will reject the null hypothesis that there is no bubble in the data.
The Neyman-Pearson lemma states that this test has the lowest probability of rejecting the null
hypothesis when the null hypothesis is false, i.e. the test has the greatest sensitivity of
detecting a bubble if one is present in the data.

We evaluate $\rho_0$ on the data realization $d$ using the fast $\Delta\chi^2$ algorithm from \S\ref{ssec:chi2}.
As remarked there, it is also computationally feasible to evaluate $\rho_0(x)$ for a
Monte Carlo ensemble of no-bubble simulations $x_0$.
To test whether $\{a_i\}=0$ is excluded at a given confidence level, we simply test whether the ``data'' value 
$\rho_0(d)$ exceeds the appropriate fraction of the simulation values $\rho_0(x_0)$.

The test statistic $\rho_0$ suffices for testing whether the data is consistent with the no-bubble model $\{a_i\}=0$.
Ideally, we would like to do more: we want to determine the full confidence regions in the $\{a_i\}$ parameter space,
rather than just being able to test whether the point $\{a_i\}=0$ is contained in a given confidence region.
We next present an algorithm for computing full confidence regions.

Conceptually, we test whether a point $\{a_i\} \ne 0$ is contained in a given frequentist confidence region
using a Monte Carlo procedure similar to the one described above for the $\{a_i\}=0$ case, but with two differences.
First, instead of the test statistic $\rho_0$, we use the statistic:
\be
\rho_{\{a_i\}}(d) = \max_{\{a_i\}'} \frac{\L(d|\{a_i\}')}{\L(d|\{a_i\})}
\ee
which is the appropriate likelihood ratio statistic for separating the model with given amplitude parameters $\{a_i\}$
from models with amplitude parameters $\{a_i\}' \ne \{a_i\}$.
Second, instead of evaluating $\rho_{\{a_i\}}$ on a Monte Carlo ensemble of no-bubble simulations, we must use an ensemble of
simulations which contain a randomly located bubble with amplitude parameters $\{a_i\}$.
This presents a computational problem: suppose we want to plot frequentist confidence regions by looping over a grid of
$\{a_i\}$ values, and testing whether each grid point is contained in a given confidence region.
Naively, this requires running a new ensemble of Monte Carlos for each grid point $\{a_i\}$, since the simulations are
$\{a_i\}$-dependent (they contain simulated bubbles with amplitude $\{a_i\}$).
Of course, this procedure is computationally impractical; we cannot afford to do an independent set of Monte Carlos
for each grid point.
However, there is a computational trick which produces a mathematically equivalent result but is computationally
affordable, allowing frequentist confidence regions to be determined.

The idea behind the trick is the following.
The reason we need a different Monte Carlo ensemble for each grid point $\{a_i\}$ is that the simulations are 
$\{a_i\}$-dependent: they contain a single bubble with amplitude parameters $\{a_i\}$.
Using notation from~\S\ref{ssec:notation}, the single-bubble simulation $x_{\{a_i\}}$ can be written:
\be
x_{\{a_i\}} = x_0 + \sum_i a_i A \beta_{Ii\n}
\ee
where $x_0$ is a no-bubble simulation.
We can imagine generating the simulation in two steps: first we randomly generate $x_0$, $I$, and $\n$,
and then we take the linear combination in the above equation to get $x_{\{a_i\}}$.
Thus, for fixed $x_0,I,\n$, the simulations with different values of $\{a_i\}$ are not fully independent; they are
all linear combinations of $(M+1)$ independent maps, where $M$ is the number of linear parameters.
A little thought shows that if we evaluate the ``data-bubble'' term from~\S\ref{ssec:chi2}
for each of these $(M+1)$ maps, this suffices to compute $\Delta\chi^2$ for an arbitrary linear combination.
This in turn suffices to evaluate the frequentist test statistic $\rho_{\{a_i\}}(x_{\{a_i\}})$ for all values of
$\{a_i\}$, which is what we need to compute confidence regions.
In other words, we can evaluate $\rho_{\{a_i\}}(x_{\{a_i\}})$ at all grid points $\{a_i\}$ simultaneously
(for a fixed choice of $x_0,I,\n$)
with computational cost proportional to $(M+1)$, not proportional to the number of grid points.

More formally, our algorithm for computing frequentist confidence regions is as follows.
\begin{enumerate}
\item At the beginning of each Monte Carlo iteration, we simulate a CMB+noise realization $x_0$ and
randomly choose a bubble radius $I$ 
(with PDF $w_I$) and center $\n$.
\item Compute the quantity
\be
Q_{I'i'\n'} = \beta_{I'i'\n'}^T A^T C^{-1} x_0
\ee
using the fast algorithm for the ``data-bubble'' term from \S\ref{ssec:chi2}.  Repeating the same algorithm for each 
$i=1,\cdots,M$, compute the quantity:
\be
R_{iI'i'\n'} = \beta_{I'i'\n'}^T A^T C^{-1} A \beta_{Ii\n}
\ee
Note that $Q_{I'i'\n'}$ is the data-bubble term for the CMB+noise realization $x_0$,
and $\sum a_{i'} R_{iI'i'\n'}$ is the data-bubble term for a realization containing
a bubble with parameters $\{ a_i, I, \n \}$, and no CMB or noise component.
\item For each grid point $\{a_i\}$, we can compute
\be
\rho_{\{a_i\}}(x_{\{a_i\}}) = \max_{\{a_i\}'} \frac{\L(x_{\{a_i\}}|\{a_i\}')}{\L(x_{\{a_i\}}|\{a_i\})}
\ee
by using the following expression to compute conditional likelihoods with $\bigoh(1)$ computational cost:
\begin{multline}
\L( x_{\{a_i\}} | \{a_i\}') \propto \sum_{I'\n'} \frac{w_{I'}}{\Npix \, (r_N-r_1)} \\
  \exp\left( \sum_{i'} a'_{i'} Q_{I'i'\n'} + \sum_{ii'} a_i a'_{i'} R_{iI'i'\n'}
               - \frac{1}{2} \sum_{i'j'} a'_{i'} a'_{j'} \beta^T_{I'i'\n'} A^T C^{-1} A \beta_{I'j'\n'} \right)
\end{multline}
Note that the last term in the exponential is the bubble-bubble term from~\S\ref{ssec:chi2},
which is the same in every Monte Carlo iteration and has been precomputed.
\item Save $\rho_{\{a_i\}}(x_{\{a_i\}})$ to disk (evaluated on a grid of $\{a_i\}$ values) and proceed to the next Monte Carlo.
\item After all Monte Carlos have been run, for each grid point $\{a_i\}$ we rank the ``data'' value
$\rho_{\{a_i\}}(d)$ relative to the ensemble of simulated values $\rho_{\{a_i\}}(x_{\{a_i\}})$ to assign a $p$-value $p(\{a_i\})$.
Confidence regions for $\{a_i\}$ can be assigned by thresholding these $p$ values, e.g.~the 95\% confidence
region consists of all values of $\{a_i\}$ satisfying $p(\{a_i\}) \leq 0.95$.
\end{enumerate}
The result of this algorithm is mathematically equivalent to running an independent set of Monte Carlos
for each grid point $\{a_i\}$.

In step 1, we constrain the random choice of bubble size $I$ and center $\n$ so that the bubble
is not obscured by the galaxy, using the same criterion~(\ref{eq:center_mask}) that
we used in the Bayesian analysis.
This constraint is necessary to make confidence regions well-defined; otherwise we would not be able
to rule out any model since we can ``hide'' an arbitrarily large bubble behind the galaxy.
Frequentist confidence regions obtained using this constraint can be rigorously interpreted as
constraints on bubbles which overlap our Hubble volume and observable sky.
This is completely analagous to the Bayesian case discussed above.

\subsection{Computational cost}

We briefly summarize the computational and storage costs of the analysis for the WMAP data.
We use 5000 Monte Carlos to calculate the bubble-bubble term and an additional
5000 Monte Carlos to compute confidence regions in the frequentist analysis.
For each Monte Carlo we must perform the $C^{-1}$ operation which takes approximately 15 core-minutes
per Monte Carlo, giving a total compute time of 2500 core-hours. We save all of the $C^{-1}$-filtered simulations
to disk and only perform the $\beta^T A^T$ operation when needed. This operation only takes a few core-seconds to compute,
but must be done for each Monte Carlo and each profile that we use.
We have over 2000 profiles and so the operation takes a few $\times \, 10^4$ core-hours in total.
At WMAP resolution each map requires approximately 6 MB of disk space (we save the maps in harmonic space
which results in a factor of 4 reduction in disk space). We require 60 GB to save all of
the $C^{-1}$-filtered maps, and an additional 13 GB to save the maps representing bubble-bubble terms.

\section{Large numbers of bubbles}
\label{sec:lots_of_bubbles}

\begin{figure}[t]
  \centering
  \includegraphics[width=10cm]{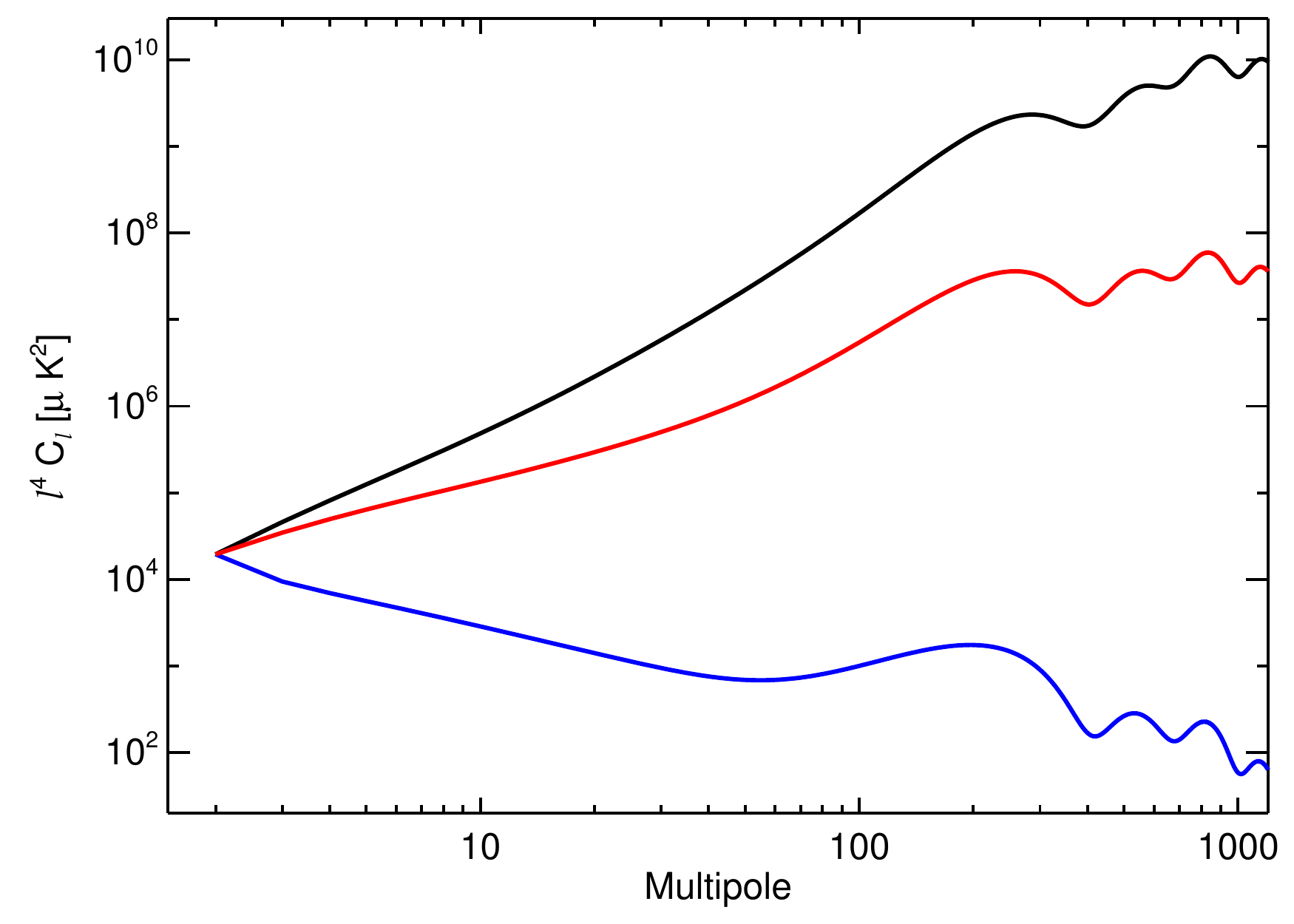}
  \caption[]{The WMAP 7-year best fit $\Lambda$CDM CMB spectrum (black line),
the spectrum of bubbles with the ramp profile normalized to the CMB spectrum at $\ell=2$ (blue),
and the spectrum of bubbles with the step profile normalized in the same way (red).}
  \label{fig:lotsofbub}
\end{figure}

In this section we consider the signal arising from a large number of overlapping bubbles.
As the number of bubble collisions increase, the signal on the sky from each additional 
collision overlaps with previous collisions.
We assume that the direction of each bubble is independent of the other bubble directions.
Then in the limit $N \gg 1$ the signal becomes a Gaussian random field by the central limit theorem.
We calculate the expected signal by summing the contributions from individual bubbles:
\be
\begin{split}
\Delta T(\n) &= \sum_{i=1}^N \Delta T_i(\n) \\
&= \sum_{\ell m} \left( \sum_{i=1}^N a_i b_{\ell}(r_i) Y_{\ell m}^*(\n_i) \right) Y_{\ell m}(\n) \label{eq:multi_bub_deltat}
\end{split}
\ee
Here, $a_i$, $r_i$ and $\n_i$ denote the amplitude, comoving displacement, and angular position
of the $i$-th bubble.  The profile $b_\ell(r)$ could be either the ramp or the step profile (for simplicity
we have not considered the case of a ramp+step model with two amplitude parameters, but this is a straightforward
generalization).
We calculate the two-point correlation function of Eq.~(\ref{eq:multi_bub_deltat}) as follows:
\be
\label{eqn:multi_bub_signal}
\left< (\Delta T)_{\ell m} (\Delta T)^{\ast}_{\ell' m'} \right>
  = \left\langle \sum_i a_i^2 b_{\ell}(r_i) b_{\ell'}(r_i) Y_{\ell m}^*(\n_i) Y_{\ell'm'}(\n_i)
    + \sum_{i\ne j} a_i a_j b_{\ell}(r_i) b_{\ell'}(r_j) Y_{\ell m}^*(\n_i) Y_{\ell'm'}(\n_j) \right\rangle
\ee
where the average $\langle\cdot\rangle$ is over realizations of the bubble random field.
The second term contains the angular average $\langle Y_{\ell m}^*(\n_i) Y_{\ell'm'}(\n_j) \rangle$.
Since the locations of bubbles $i\ne j$ are assumed independent, this is equal to 
$\langle Y_{\ell m}^*(\n_i) \rangle \langle Y_{\ell'm'}(\n_j) \rangle = \delta_{\ell 0} \delta_{\ell' 0} / (4\pi)^2$
and only contributes a monopole.  We therefore ignore the second term in Eq.~(\ref{eqn:multi_bub_signal})
because a bubble contribution to the monopole is not measurable (since $T_{\rm CMB}$ is a free parameter anyway).

We can evaluate the first term in Eq.~(\ref{eqn:multi_bub_signal}) as follows.
The parameters $a_i$, $r_i$ and $\n_i$ are assumed indepdendent, so we can average them separately.
The average over $r_i$ can be performed using the uniform prior (Eq.~(\ref{eq:size_prior})):
\be
\left\langle b_{\ell}(r_i) b_{\ell'}(r_i) \right\rangle = \frac{1}{r_{\rm max}} \int_0^{r_{\rm max}} dr\, b_{\ell}(r) b_{\ell'}(r)
\ee
The average over angular position $\n_i$ is straightforward to compute:
\be
\left\langle Y_{\ell m}^*(\n_i) Y_{\ell'm'}(\n_i) \right\rangle = \frac{1}{4\pi} \delta_{\ell\ell'} \delta_{mm'}
\ee
Plugging into Eq.~(\ref{eqn:multi_bub_signal}), we can now read off the power spectrum of the bubble
contribution to the CMB temperature as follows:
\be
\label{eqn:multi_bub_result}
C^{\rm bub}_\ell = \frac{N \left<a^2\right>}{4\pi} \frac{1}{r_{\rm max}} \int_0^{r_{\rm max}} dr\, b_{\ell}(r)^2
\ee
where $\langle a^2 \rangle = \langle a_i^2 \rangle$ and the expected number of bubbles $N$ in
the observable range $-r_{\rm max} \le r \le r_{\rm max}$, are model-dependent quantities.
We see from this expression that
the combination $N \langle a^2 \rangle$ is constrained by the power spectrum.

The many-bubble power spectrum has also been calculated recently by Aguirre \& Kozaczuk~\cite{Kozaczuk:2012sx}.
Our result agrees if we use the ramp profile and ignore transfer functions, i.e.~make the Sachs-Wolfe approximation in Eq.~(\ref{eq:sw_ramp}).
For the ramp profile, we find that the Sachs-Wolfe approximation is fairly accurate but overpredicts $C_\ell^{\rm bub}$
by $\approx 14$\%, due to omitting ISW contributions to the bubble profile as described in~\S\ref{ssec:cmb_transfer}.
We note that in~\cite{Kozaczuk:2012sx}, higher-order corrections to the linear CMB + bubble model are
calculated, but shown to be negligible.

In Fig.~\ref{fig:lotsofbub}, we show the power spectrum $C_\ell^{\rm bub}$
for both the ramp and step models.
We use the denser sampling of $r$ that we used for the step profile
in order to calculate the integral more precisely.
The power spectrum falls off roughly as $1/\ell^5$ for the ramp profile,
and $1/\ell^3$ for the step profile.
Given the roughly $1/\ell^2$ dependence of the CMB spectrum, this implies
that the step model is constrained by a wide range of $\ell$ values, while
the ramp model is constrained mainly by the lowest values of $\ell$
(in fact we find that 92\% of the statistical weight comes from the quadrupole).
For the ramp model, the Sachs-Wolfe approximation is fairly accurate and Planck will not
significantly improve WMAP constraints; the opposite statements are true for the step model.
This parallels the discussion in~\S\ref{ssec:cmb_transfer} above.

\section{Conclusions}
\label{sec:conc}

Searching for anomalous signals in CMB data has become a big industry, with a large number of
different methodologies being employed~\cite{Bennett:2010jb}.
Signals that may be in the CMB data offer the opportunity to learn about the inflationary,
and possibly the pre-inflationary, epoch.
In this paper, we have focused on an example signal that has recently been
discussed in the theory literature.
This theory states that we live in a
bubble of low vacuum energy density surrounded by an infinite, eternally inflating spacetime
with a higher vacuum energy density.
If there are other bubble regions that were created near to our own then they could have collided
with our bubble in the past leaving a distinct pattern in the CMB.

We have developed a toolkit of algorithms which allow us to perform
the exact, all-sky, optimal data analysis for the bubble signal.
The main features of our analysis are:
\begin{itemize}
\item We precompute the bubble profiles including CMB transfer functions.
\item We precompute a large number of maps (the maps denoted $D_{Ii}(\n)$ and $\langle X_{Ii}(\n) X_{Ij}(\n) \rangle$
in~\S\ref{ssec:chi2}) which permit very fast evaluation of $\Delta\chi^2$, essentially as a table lookup operation.
\item After these precomputations, the Bayesian posterior likelihood can be evaluated so quickly that a
Bayesian data analysis may be performed trivially by gridding the likelihood (i.e.~without MCMC).  We compute
the exact all-sky posterior without making any approximations, such as neglecting small-scale anisotropy in the noise.
\item We identify the optimal frequentist statistic for the bubble collision problem, and show how a full frequentist analysis, including
calculation of confidence regions, can be made computationally affordable with some additional computational tricks.
\end{itemize}
Although we have focused on the bubble collision problem, our algorithms should apply to any parametrized
family of azimuthally symmetric profiles, and we have presented them in a form which emphasizes this generality.
Our algorithms should be useful for other problems, e.g.~searches for other defects such as textures, or
optimal detection of Sunyaev-Zeldovich clusters.
We analyze WMAP data in a separate paper~\cite{short}.

We have also calculated the signal that would be expected from a large number of independent bubbles.
The signals from different bubbles overlap, and the total signal tends to a Gaussian random field as $N \gg 1$, which
implies that the power spectrum is the optimal statistic in the many-bubble limit.
We calculate the many-bubble power spectrum and find results consistent with~\cite{Kozaczuk:2012sx}.
Our methodology could be extended to the case of a small number of collisions.
This is slightly different to the single-bubble analysis because the bubbles can overlap.
The extra parameters for the additional bubbles can be included in the likelihood, which
can be calculated in a similar way to the single-bubble case, albeit with increased computational
requirements.  We defer the details to future work.

Future observations can improve WMAP constraints on bubble collisions in a few ways.
Considering CMB observations first, the increased angular resolution of upcoming experiments
such as Planck will improve constraints on the step model, but not on the ramp model where
the signal is weighted toward large scales that are already sample variance limited in WMAP.
Future measurements of CMB polarization may eventually improve constraints on both models~\cite{Czech:2010rg}.
Looking beyond the CMB, large-scale structure can potentially provide interesting
constraints on bubble collisions, e.g.~coherent galaxy flows~\cite{Larjo:2009mt,Ahn:2012fh},
Lyman-alpha forest measurements~\cite{Weinberg:2003eg}, or 21-cm radiation
surveys~\cite{Loeb:2003ya,Kleban:2007jd}.

\section*{Acknowledgments}

We thank Adam Brown, Matt Kleban, Ben Freivogel, Steve Shenker, and Lenny Susskind for helpful discussions.
SJO acknowledges support from the US Planck Project, which is funded by the NASA Science Mission
Directorate.
LS~is supported by DOE Early Career Award DE-FG02-12ER41854 and the National Science Foundation
under PHY-1068380.
KMS was supported by a Lyman Spitzer fellowship in the Department of Astrophysical Sciences at Princeton University.
Research at Perimeter Institute is supported by the Government of Canada
through Industry Canada and by the Province of Ontario through the Ministry of Research \& Innovation.
Some of the results in this paper have been derived using the HEALPix~\cite{Gorski:2004by} package.
Computations were performed at the TIGRESS high performance computer center at Princeton University which is jointly 
supported by the Princeton Institute for Computational Science and Engineering and the Princeton University Office of 
Information Technology.
We acknowledge the use of the Legacy Archive for Microwave Background Data Analysis (LAMBDA), part of the High Energy Astrophysics Science Archive Center (HEASARC). 
HEASARC/LAMBDA is a service of the Astrophysics Science Division at the NASA Goddard Space Flight Center.


\begingroup\raggedright\endgroup

\appendix

\section{Details of the $C^{-1}$ filter}
\label{app:c_inverse}

In this section we overview the calculation of the inverse-variance filtering operation $x \rightarrow C^{-1} x$, 
where, following notation introduced in~\S\ref{ssec:notation}, $x_{\mu p}$ is a per-channel pixel-space map
and $C = N + A S A^T$.
Additional details can be found in~\cite{Smith:2007rg} where the method that we describe was first developed.

For purposes of this paper, it suffices to implement the filtering operation $x \rightarrow A^T C^{-1} x$
where the quantity on the RHS is a harmonic-space map.
(Throughout the body of the paper, $C^{-1}$ only appears as part of the combination $A^T C^{-1}$.)
Now consider the identity:
\begin{eqnarray}
A^T C^{-1} &=& A^T (N + A S A^T)^{-1} \nn \\
           &=& S^{-1/2} (1 + S^{1/2} A^T N^{-1} A S^{1/2})^{-1} S^{1/2} A^T N^{-1}   \label{eq:cinv_filter_rearranged}
\end{eqnarray}
This identity is not obvious, but can be proved by mutiplying both sides on the right by $(N + A S A^T)$.
It will be convenient to use the RHS of Eq.~(\ref{eq:cinv_filter_rearranged}), so that the filtering operation
can be performed using the purely harmonic-space operation $a \rightarrow X^{-1} a$, where we have defined the
operator:
\be
X = 1 + S^{1/2} A^T N^{-1} A S^{1/2}
\ee
A further advantage of using the RHS of Eq.~(\ref{eq:cinv_filter_rearranged})
is that the inverse noise covariance $N^{-1}$ appears instead of the noise covariance $N$.
This allows us to incorporate a foreground mask, by setting matrix entries of $N^{-1}$ to
zero in masked pixels.  Analagously, we marginalize the CMB monopole and dipole (independently
in each channel $\mu$) by modifying the operator $N^{-1}$ so that $N^{-1}x = 0$ for the four
independent modes of the monopole or dipole.  (This implies that $C^{-1}x$ is also zero for
these modes.)

We implement the filtering operation $a \rightarrow X^{-1}a$ using the preconditioned
conjugate gradient algorithm~\cite{PCGpaper}.
The preconditioner must be chosen with care
since it has a significant impact on the compute time. For example,
a diagonal preconditioner gives extremely slow convergence of the algorithm
because on scales $\ell \lesssim 500$ the data is signal
dominated~\cite{Larson:2010gs,Komatsu:2010fb} and so the approximation that the covariance is diagonal
is not valid.
Instead, a block diagonal preconditioner $P$ can be used that contains the full covariance
on large scales $\ell < \ell_{\rm split}$ and is diagonal on smaller scales:
\be
\label{eqn:precon1}
P = \left( \begin{array}{cc}
X^{-1}_{(0)} & 0 \\
0 & X^{-1}_{\Delta} \end{array} \right)
\ee
where the subscript $(0)$ denotes that the full resolution is used
and $X^{-1}_{\Delta}$ is the diagonal approximation.
The block diagonal preconditioner is found to have faster convergence than the diagonal
preconditioner~\cite{Smith:2007rg}, but is still too slow for our purposes.
A multigrid preconditioner is found to decrease the convergence time by an order of
magnitude over the block diagonal preconditioner for the WMAP data~\cite{Smith:2007rg}.
The aim of the multigrid preconditioner is to perform the conjugate gradient descent at resolution
$N_{\rm side}$ with a preconditioner that is itself calculated using the conjugate gradient descent
algorithm but at a lower resolution of $N_{\rm side}/2$.
The $X_{(0)}^{-1}$ operation is therefore evaluated using conjugate gradient descent
with a preconditioner $X_{(1)}^{-1}$ that is calculated at a lower resolution
such that $l_{\rm max}^{(1)} < l_{\rm max}^{(0)}$.
The lower resolution preconditioner can itself
be calculated using an even lower resolution preconditioner $X_{(2)}^{-1}$ in a recursive manner.
The preconditioner for the first layer of recursion is:
\be
P = \left( \begin{array}{cc}
X^{-1}_{(1)} & 0 \\
0 & X^{-1}_{\Delta} \end{array} \right)
\ee
where the subscript $(1)$ denotes that the preconditioner is calculated at resolution $N_{\rm side}/2$.
The time required to calculate harmonic transforms depends on the resolution as
$\mathcal{O}\left(N_{\rm side}^3\right)$, and so operations at the coarser
resolution are significantly faster.
The inversion therefore proceeds recursively with the coarsest resolution of $N_{\rm side} = 128$
preconditioned using the block diagonal preconditioner~\cite{Hirata:2004rp,Eriksen:2004ss}.
Using the multigrid preconditioner
we find that $C^{-1} x$ can be calculated in approximately 15 core-minutes, which together with
the other computational tricks that we describe makes our likelihood analysis computationally feasible.

\section{Likelihood model}
\label{app:like_model}

\begin{figure}[t]
  \centering
  \includegraphics[width=8.5cm]{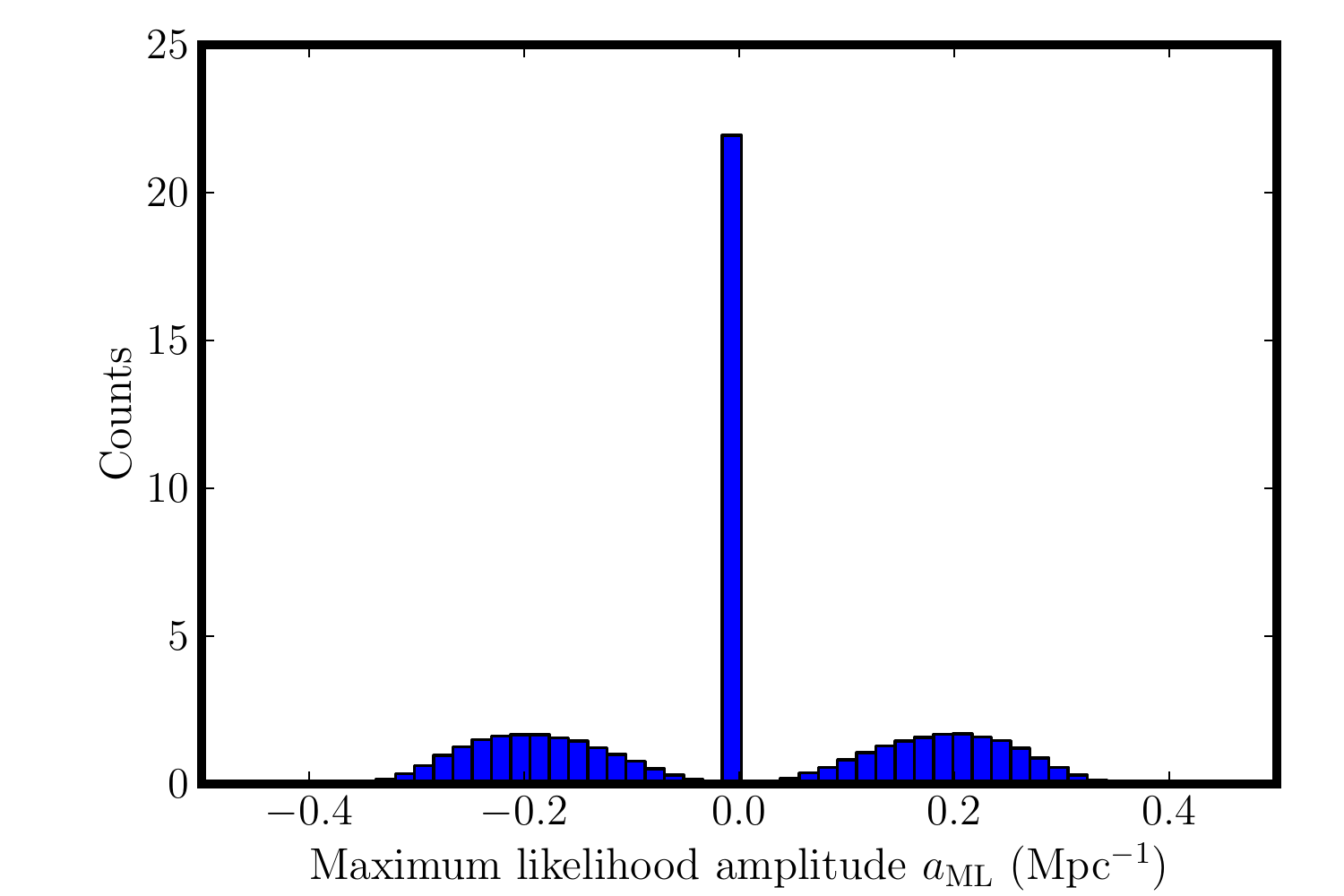}
  \includegraphics[width=8.5cm]{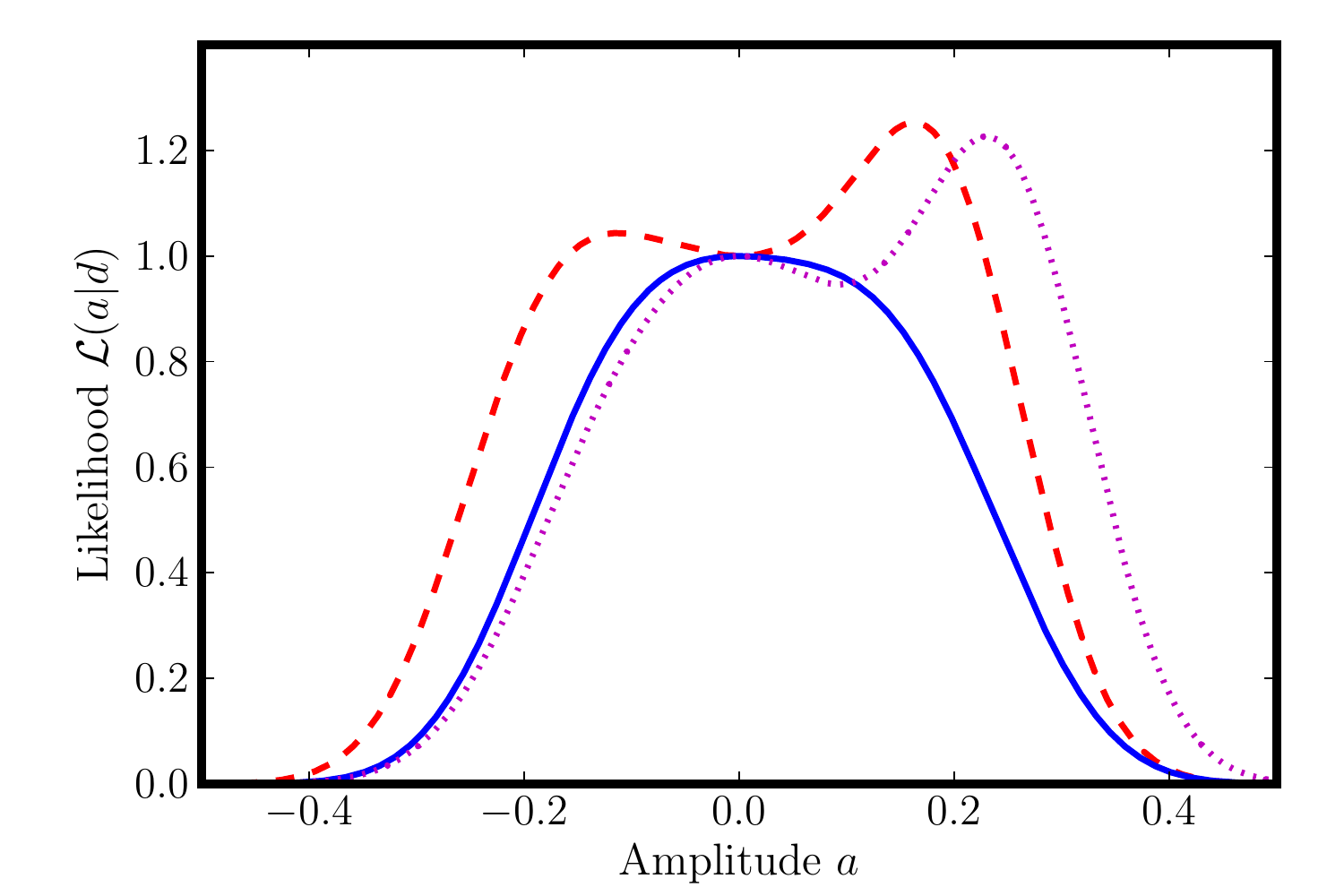}
  \caption[]{{\em Left panel:} Histogrammed maximum likelihood bubble amplitude $a_{\rm ML}$,
     for many Monte Carlo simulations of the data $d$.
     Counterintuitively, $a_{\rm ML}$ is within $0.001\sigma$ of zero 11\% of the time.
  {\em Right panel:} Posterior likelihoods $\L(d|a)$ for three randomly simulated data realizations $d$.
    It is seen that the derivative $(\partial\L/\partial a)$ is always zero at $a=0$, but this point can be either
  the global maximum likelihood (blue, solid curve), a local maximum which is not the global maximum (magneta, dotted curve),
  or a local minimum (red, dashed curve).
    Likelihoods in this figure were generated using a toy model with 1000 $\mu$K-arcmin isotropic noise, 
  $5^\circ$ Gaussian beam, no sky cut, and assuming fixed bubble size $\thetabubble = 30^\circ$.}
  \label{fig:toy_likelihood}
\end{figure}

The posterior likelihood $\L(\{a_i\}|d)$ defined in Eq.~(\ref{eq:posterior}) is a
complicated function of the data that cannot be easily evaluated analytically. However, by making some
approximations we can create a model of the likelihood that can be used to understand its main properties.
In particular we wish to understand the following phenomenon.
Suppose that we make Monte Carlo simulations of the data $d$, and compute the value $a_{\rm ML}$ of the
amplitude parameter which maximizes the likelihood $\L(a|d)$ in each simulation.
Naively we might guess that $a_{\rm ML}$, histogrammed over many Monte Carlo
simulations, would be Gaussian distributed with mean zero and width roughly equal
to the minimum detectable bubble amplitude.
Instead, $a_{\rm ML}$ is distributed as shown in the left panel of Fig.~\ref{fig:toy_likelihood}.
There is an order-one probability for $a_{\rm ML}$ to be very close to zero.
While this phenomenon may seem counterintuitive, we will show that it has a natural explanation
in terms of analytic properties of the likelihood.

A clue can be obtained by visually inspecting the likelihood
$\L(a|d)$ for a few randomly generated data realizations $d$ (Fig.~\ref{fig:toy_likelihood},
right panel).  It is seen that the derivative $\partial\L/\partial a$ is always zero at
$a=0$, but this local extremum can be either a local maximum or a local minimum.
These properties of the likelihood can be understood analytically as follows.

For notational simplicity, consider the simplest case of a single linear parameter $i$,
and non-linear parameters $I$ which can only take a single value, i.e.~a randomly located
profile with arbitrary amplitude but fixed radius and shape.
Suppressing superfluous indices, the posterior likelihood can be written:
\be
\mathcal{L}(a|d) = \frac{1}{\Npix} \sum_{\n} 
  \exp{\left( a \; \beta_{\n}^T A^T C^{-1} d - \frac{a^2}{2} \beta_{\n}^T A^T C^{-1} A \beta_{\n} \right)}
\label{eq:toy_likelihood}
\ee
The first derivative at $a=0$ is:
\be
\left( \frac{\partial \mathcal{L}(a|d)}{\partial a} \right)_{a=0} 
  = \frac{1}{\Npix} \sum_{\n} \beta_{\n}^T A^T C^{-1} d  \label{eq:first_derivative}
\ee
In the limit $\Npix\rightarrow \infty$ of a finely pixelized map, the RHS approaches zero.
To see this, we first take the $\Npix\rightarrow\infty$ limit of the harmonic-space map
$\Npix^{-1} \sum_{\n} \beta_{\n}$:
\begin{eqnarray}
\frac{1}{\Npix} \sum_{\n} (\beta_{\n})_{\ell m} 
  & \rightarrow & \frac{1}{4\pi} \int d^2\n\, (\beta_{\n})_{\ell m}  \nn \\
  & = & \frac{1}{4\pi} \int d^2\n\, b_\ell Y_{\ell m}^*(\n) \nn \\
  & = & \frac{b_\ell }{\sqrt{4\pi}} \delta_{\ell 0}
\end{eqnarray}
i.e.~the map $\Npix^{-1} \sum_{\n} \beta_{\n}$ is a pure monopole.  When we apply $C^{-1} A$ to this map,
we get zero since there is no instrumental sensitivity to the monopole.  (More formally, as explained in
Appendix~\ref{app:c_inverse}, the monopole has been marginalized by assigning it infinite noise variance
in the pixel domain, which implies that we get zero when we multiply by $C^{-1}$.)
This implies that the RHS of Eq.~(\ref{eq:first_derivative}) is zero.\footnote{Strictly speaking, this is
only true in the high resolution limit $\Npix\rightarrow \infty$.
In a finite pixelization, the RHS will be nonzero due to pixelization artifacts, but small (of order $1/\Npix$).
Our procedure for removing terms in the likelihood which
correspond to bubbles which are hidden behind the galactic mask, 
given in Eq.~(\ref{eq:center_mask}) above,
also gives a small nonzero contribution to the
derivative on the RHS of Eq.~(\ref{eq:first_derivative}).
These effects mean that the local extremum of the likelihood does not appear precisely at $a=0$, but is perturbed
to a value of $a$ which is nonzero but very small.}

The second derivative of Eq.~(\ref{eq:toy_likelihood}) at $a=0$ is:
\be
\left( \frac{\partial^2 \mathcal{L}(a|d)}{\partial a^2} \right)_{a=0} 
   = \frac{1}{\Npix} \sum_{\n} (\beta_{\n}^T A^T C^{-1} d)^2 - \beta_{\n}^T A^T C^{-1} A \beta_{\n}  \label{eq:second_derivative}
\ee
Here, the RHS is nonzero on a per-realization basis.
However, using the expectation value $\langle d d^T \rangle = C$, one sees that the
expectation value of Eq.~(\ref{eq:second_derivative}) is zero.
This implies that the second derivative will be positive (or negative) some order-one fraction of the time,
in order to get a zero expectation value.

Summarizing, we have now shown that the likelihood ${\mathcal L}(a|d)$ always has a local extremum at $a=0$,
which is a local maximum an order-one fraction of the time.  If $a=0$ is a local maximum, then there is a further
order-one probability that it turns out to be the global maximum, which can be seen intuitively because
the likelihood has a finite number
of local maxima which are candidates for the global maximum.
This explains the phenomenon
seen in the histogram in the left panel of Fig.~\ref{fig:toy_likelihood} above, where the maximum
likelihood is at $a=0$ an order-one fraction of the time.

How does this phenomenon affect our statistical analysis?
We answer this question separately for the Bayesian (\S\ref{ssec:bayesian}) 
and frequentist (\S\ref{ssec:frequentist}) cases.
In the Bayesian case, we infer confidence regions on the amplitude parameters $a_i$ from the posterior
likelihood $\L(a_i|d)$, using the prescription in 
Eqs.~(\ref{eq:bayesian_confidence_regions1}),~(\ref{eq:bayesian_confidence_regions2}).
In a data realization where $\L(a_i|d)$ has its global maximum at $a_i=0$, we will find that the
point $a=0$ is contained in every confidence region, i.e.~if instead of computing the 95\% CL region,
we use a different $p$-value, we will find that the confidence region contains $a_i=0$, irrespective
of the $p$-value.

The frequentist case is similar:
if we have a realization $d$ where $\L(a_i|d)$ has its global maximum near $a_i=0$,
we will find that the frequentist test statistic $\rho_0(d)$ is nearly equal to 1.
It follows that $a=0$ is contained in every frequentist confidence region, regardless
of the threshold (i.e.~whether 68\%, 95\%, etc.)
This is because we determine whether $a=0$ is outside the confidence
region by testing whether $\rho_0(d)$ is anomalously {\em high} compared to an ensemble
of simulated values $\rho_0(d')$, and the simulated values satisfy $\rho_0(d') \ge 1$ by definition.
Therefore, in either the Bayesian or frequentist framework, 
there is an order-one probability that a random data realization $d$
will have a maximum likelihood amplitude precisely at $a_i=0$, which
implies that $a=0$ is contained in every confidence region.
This behavior is expected analytically and
simply means that the realization is consistent with being a Gaussian field, and there is no
statistical evidence for a profile with nonzero amplitude.

\end{document}